\documentclass[aoas,preprint]{imsart}

\RequirePackage[OT1]{fontenc}
\RequirePackage{amsthm,amsmath, amssymb}
\RequirePackage{color, subfigure, graphicx, comment, bm, url}
\RequirePackage{natbib}
\RequirePackage[colorlinks,citecolor=blue,urlcolor=blue]{hyperref}

\arxiv{math.PR/0000000}

\DeclareRobustCommand{\vect}[1]{\bm{#1}}   
\newcommand{\TR}[1]{\textcolor{black}{#1}}

\begin{document}
\begin{frontmatter}
\title{Bayesian Non-Homogeneous Markov Models via Polya-Gamma Data Augmentation with Applications to Rainfall Modeling}

\begin{aug}
\author{\fnms{Tracy} \snm{Holsclaw}\thanksref{m1}\ead[label=e1]{tholscla@ams.ucsc.edu}},
\author{\fnms{Arthur M.} \snm{Greene}\thanksref{m2}\ead[label=e2]{amg@iri.columbia.edu}},
\author{\fnms{Andrew W.} \snm{Robertson}\thanksref{m2}\ead[label=e3]{awr@iri.columbia.edu}}
\and
\author{\fnms{Padhraic} \snm{Smyth}\thanksref{m1}\ead[label=e4]{smyth@ics.uci.edu}}

\affiliation{University of California, Irvine\thanksmark{m1} and Columbia University\thanksmark{m2}}

\address{Department of Computer Science\\
University of California, Irvine\\
Irvine, CA 92697\\
\printead{e1}\\
\phantom{E-mail:\ }\printead*{e4}}

\address{International Research Institute for Climate and Society\\
 The Earth Institute at Columbia University\\
230 Monell, 61 Route 9W - PO Box 1000\\
 Palisades, New York 10964-8000\\
\printead{e2}\\
\phantom{E-mail:\ }\printead*{e3}}
\end{aug}

\begin{abstract}   
Discrete-time hidden Markov models are a broadly useful class of latent-variable models  with applications in areas such as speech recognition, bioinformatics, and climate data analysis.  It is common in practice to introduce temporal non-homogeneity into such models by making the transition probabilities dependent on time-varying exogenous input variables via a multinomial logistic parametrization. We extend such models to introduce additional non-homogeneity into the emission distribution using a generalized linear model (GLM), with data augmentation for sampling-based inference. However, the presence of the logistic function in the state transition model significantly complicates parameter inference for the overall model, particularly in a Bayesian context. To address this we extend the recently-proposed Polya-Gamma data augmentation approach to handle non-homogeneous hidden Markov models (NHMMs), allowing the development of an efficient Markov chain Monte Carlo (MCMC) sampling scheme. We apply our model and inference scheme to 30 years of daily rainfall in India, leading to a number of insights into rainfall-related phenomena in the region. Our proposed approach allows for fully Bayesian analysis of relatively complex NHMMs on a scale that was not possible with previous methods.  Software implementing the methods described in the paper is available via the R package NHMM.
\end{abstract}

\begin{keyword}
\kwd{non-homogenous hidden Markov model}
\kwd{multivariate time series}
\kwd{Polya-Gamma latent variables}
\kwd{probit and logit link}
\end{keyword}

\end{frontmatter}

\section{Introduction}  

Consider the problem of modeling the dynamics of a multivariate discrete time series ${\bm y}_t$,  with component measurements $y_{ts}$,  $s  = 1,\ldots,S$, and   a discrete-time index $t=1,\ldots,T$. A useful modeling approach in this context is the hidden Markov model (HMM), where the  observed  ${\bm y}_t$'s are assumed to be a stochastic function of a (hidden) finite-state Markov  process ${\bm z}$, with components $z_t \in \{1,...,K\}$, and  where each vector $ \bm{y}_t$ is assumed to be conditionally independent of all  other $ \bm{y}_{t'}$ vectors and state variables $z_{t'}$, $t' \ne t$, given  state $z_t$ \citep{langrock2}. The conditional distribution of the ${\bm y}_t $ vectors at time $t$ given the state $z_t$ is often  assumed to be time-homogeneous, defined by so-called emission distributions,\footnote{Here we use the term ``emission distributions'', widely used in speech recognition and language modeling (e.g., \cite{emission}) - these are also referred to as ``state-dependent distributions'' (e.g., \cite{langrock2})} $f({\bm y}_t| z_t = k,{\bm \theta}), k \in \{1,\ldots,K\}$ where ${\bm \theta}$ represents the emission distribution parameters. The distributional choice for $f$ will depend on the particular characteristics of the ${\bm y}_t$ measurements for a given application.   

 HMMs are appealing for problems where  the dynamics of ${\bm y}_t$ are too complex to be directly modeled (e.g., for high-dimensional problems where $S$ is large) but can instead be approximated  via a discrete-state hidden Markov process ${\bm z}$. For example, a common assumption in practice (and one that is used in this paper - see also \cite{langrock2}, p.140, and the discussion of contemporaneous conditional independence), is to assume that the components of ${\bm y}_t$ are conditionally independent given the state, i.e., that $f({\bm y}_t| z_t = k,{\bm \theta}) = \prod_{s=1}^S f_s(y_{ts} | z_t = k,{\bm \theta})$, where $f_s(\cdot)$ denotes the conditional distribution of component $s$ of the observed vector ${\bm y}_t$. HMMs can also be used to produce a time-dependent clustering of the observations ${\bm y}_t$, where the state variables $z_{t}$ are interpreted as indicators of cluster memberships, with the Markov dependence providing temporal dependence (in contrast to mixture model clustering for example, where the cluster memberships are modeled as being independent). Using HMMs for clustering in this manner can be useful in econometric, ecological, or other scientific time-series applications (e.g., \cite{macd,raphael1999automatic, siepel2004combining, mamon2007hidden, langrock1}). The goal   is often to try to gain insight into possible latent processes that might be giving rise to the observed ${\bm y}_t$ data, for example by analyzing and interpreting  differences among the emission distributions  $f({\bm y}_t | z_t = k,{\bm \theta})$ across states.

The time-homogeneity of the standard HMM (at the parameter level, as described above)  can be limiting in practice, for example if  ${\bm y}_t$ has seasonal dependence or is non-stationary. One approach to relaxing this assumption is to  allow the $K \times K$ transition matrix  probabilities to  be dependent on an exogenous time-series ${\bm x}_t$, resulting in a non-homogeneous hidden Markov model (NHMM) (e.g., \cite{hughes94,diebold,hughes1999non,Kirsh04,kim2008estimation,Paroli08,meligkotsidou2011forecasting,raja2}). A natural parametrization is to model each of the $K$ rows of the transition matrix via a multinomial logistic function, with $K$ possible outcomes (the $K$ possible states at time $t+1$ given the current state $z_t$). Temporal inhomogeneity can also be introduced in the emission component of the model, for example by allowing the parameters of the emission distributions $f({\bm y}_t | z_t = k,{\bm \theta})$ to vary with time $t$ and location $s$ as a function of another exogenous process ${\bm w}_{ts}$ (e.g., \cite{holsclaw2015bayesian}) .

\begin{figure}[htp]
\begin{center}
\includegraphics[scale=0.4]{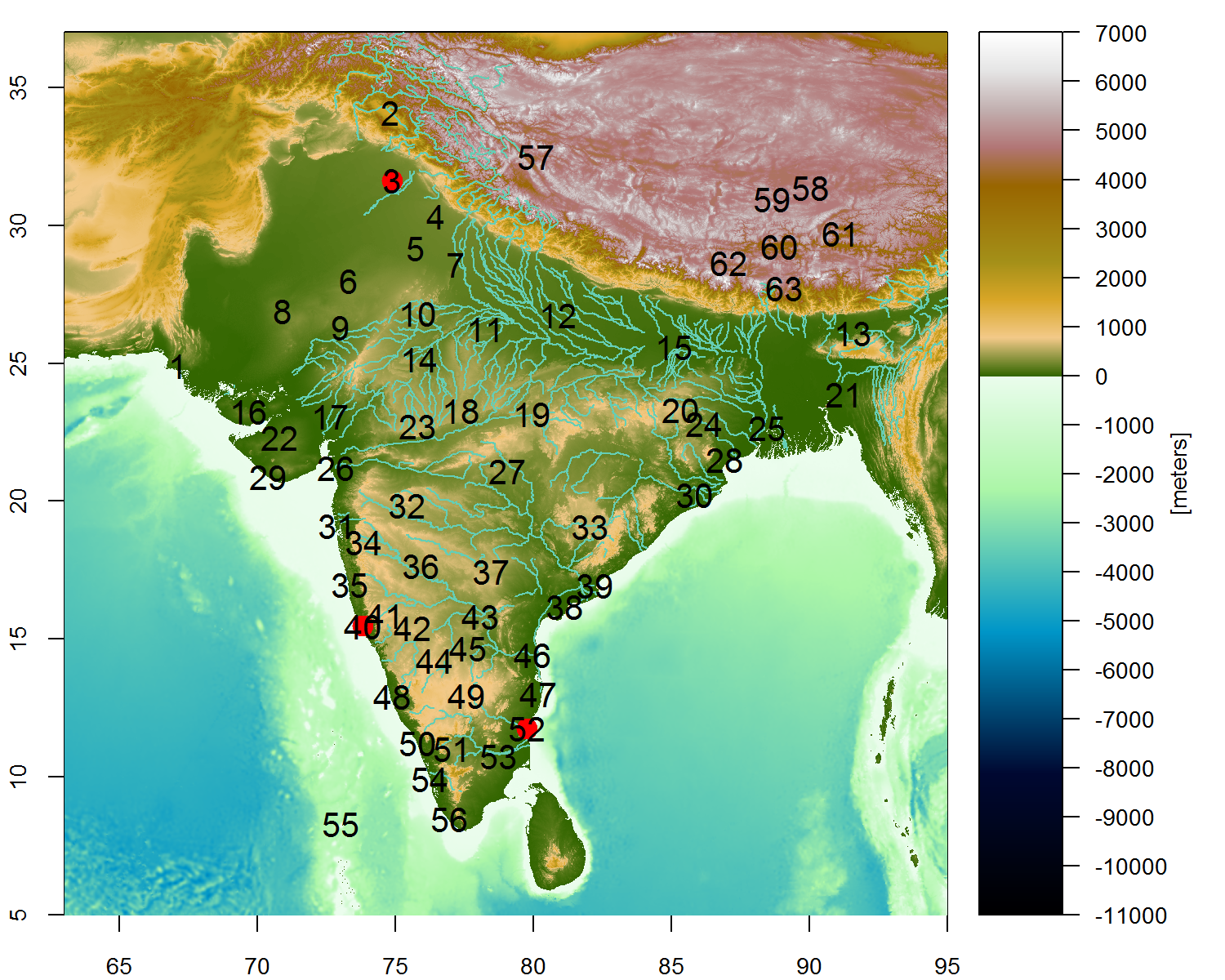}  
\end{center}
\caption{Locations of the 63 rain gauge stations, showing the topography of South Asia. Stations 3  (31.63$^\circ$N,74.87$^\circ$E), 40 (15.48$^\circ$N,73.82$^\circ$E), and 52  (11.77$^\circ$N,79.77$^\circ$E) are each marked with a dot; these diverse locations will be used in subsequent plots as examples.}
  \label{statloc}
\end{figure}

\begin{figure}[htp]
\begin{center}
\includegraphics[scale=0.60]{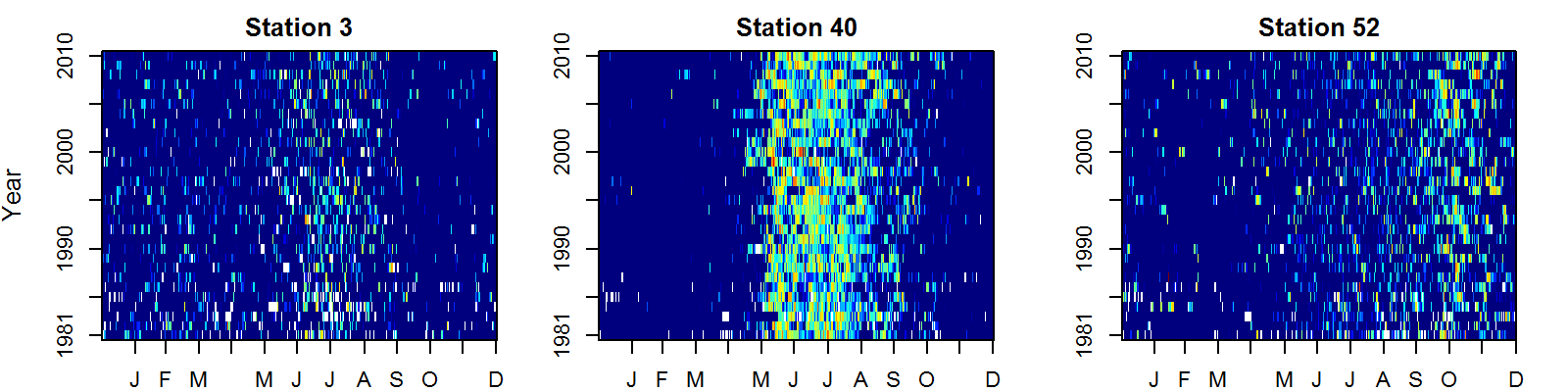}
\end{center}
\caption{Daily rainfall data (log of the amount in mm) with the x-axis being the day of the year and the y-axis depicting the 30 years. The left panel shows, the relatively dry, station 3 located in NW India, the middle panel shows station 40 on the west coast, strongly impacted by the summer monsoon,  and the right panel showing station 52 on the SW coast that is influenced by the winter monsoon, peaking in October--December. Darker colors indicates lower daily log rainfall amounts and lighter colors indicates higher daily values; white is for missing observations.}
  \label{raws}
\end{figure}

As a motivating example we consider the problem of modeling and
simulating daily station rainfall data over India where the
observations $y_{t,s}$ correspond to the amount of rain that has
fallen on day $t$ at weather station $s$. The data we analyze has been
collected daily for 30 years at 63 rain gauge stations across India,
totaling well over half a million observations ($6.9 \times 10^5$).
The geographical area of interest contains diverse sub-regions where
the rainfall varies greatly in seasonal timing and amount. As shown in
Figure \ref{statloc} some stations lie in the Himalayas while
  others are located variously in coastal, monsoon or desert regions; these data are not isotropic in nature. Figure
\ref{raws} shows the rainy days in lighter shades indicating amounts
and dry days in dark shades for three contrasting stations (see
\cite{supplementA} for additional stations).

Accurately modeling and simulating rainfall on a daily time scale is
important across a number of diverse applications such as crop
modeling, flood risk assessment, and water policy decisions
\citep{HansenEtAl_CR06, challinor2009crops,piani2010statistical}.
Multivariate HMMs have been successfully applied to this modeling
problem in the past, where the hidden variables $z_t$ can be
interpreted as weather states exhibiting persistence at daily
time scales, and the emission distributions $f({\bm y}_t | z_t = k,{\bm \theta})$
capture the spatial and distributional characteristics of observed
rainfall for each state
\citep{zucchiniguttorp91,hughes94,kirshner05,greene08, langrock2}.  Of direct
interest to climatologists is the situation where the rainfall in a
given region is being influenced or driven by time-varying atmospheric
variables ${\bm x}_t$ such as pressure differentials at large spatial
scales.  Relating these large scale variables to local rainfall
characteristics at particular station locations $s$ is known as {\it
  downscaling}. NHMMs have been found to be broadly useful in this
context where the ${\bm x}_t$ variables act as ``drivers" for the
Markov transition matrix as described earlier \citep{hughes1999non,
  bellone00,charles2004statistical, robertson09, germain10,
  careysmith, heaps15}. Other work in a downscaling context, such as that of \cite{berrocal} and \cite{fuentes} for ozone and airborne particulates, focuses on the use of Gaussian models - these models are not appropriate here given the non-negativity and non-Normality of precipitation data.

There are a multitude of other modeling approaches that could be used in this context. In particular, dynamic spatio-temporal models provide a rich framework for modeling spatial and temporal dependencies. These models often use continuous latent-space representation (in contrast to the discrete state representation of the HMM approach) and are often parametrized in a manner that can incorporate relevant scientific knowledge, e.g., in the form of differential equations (see \cite{hooten}, for a review). Such approaches can provide richer representations for spatial structure that go beyond the conditional independence assumption that has often been used when NHMMs are applied to precipitation modeling (and that we use here in this paper). In Section \ref{spatialsec} and in \cite{supplementA} we examine the model’s ability to capture spatial dependence across stations and conclude that while the conditional independence approach tends to underestimate the true spatial dependence, that the model nonetheless is capturing much of the dependence that is empirically observed. For applications where spatial dependence is of critical importance, additional spatial dependence could be incorporated in the emission component of our proposed model at the cost of additional complexity and computational effort.

Many of the early applications of HMMs and NHMMs, to climate data as well as to other problems, have relied on point estimates of model parameters, often using the Expectation-Maximization (EM) algorithm  for parameter estimation \citep{dempster77}. There is, however, a growing need for efficient Bayesian methods  for assessing uncertainty in these types of models (e.g., \cite{ryden08}). For example, in the context of climate data, modeling the  uncertainty in rainfall amounts is important in both seasonal forecasting and climate change downscaling applications \citep{maraun2010precipitation,vermeulen2013addressing} and Bayesian simulations are better suited to characterizing such uncertainty than point-estimate approaches.  

While there has been extensive development of Bayesian methods for HMMs \citep{scott02,fruhwirth2006finite,ryden08,langrock1} there has been little work on Bayesian estimation of NHMMs. Prior work has typically focused on analysis of  small univariate data sets due to the complexity and computational expense of the Metropolis-Hastings MCMC schemes used for inference (e.g., \cite{filardo1998business,spezia2014modelling}).  \cite{meligkotsidou2011forecasting} apply the Bayesian multinomial logit regression (MNL) latent variable technique developed by \cite{Holmes1,Holmes2} to the NHMM, illustrating the approach using a relatively small univariate financial econometrics data set with monthly observations over 38 years. For many applications however we need methods that scale up efficiently to much larger data sets. The rainfall data set we analyze later in the paper consists of a 63-dimensional time-series with $T \approx 30 \times 365 = 10950$ observations per time-series.

The development of an efficient Bayesian sampling scheme to handle
logistic transition matrices in NHMMs, a problem that has proven
challenging in the past because of the lack of conjugacy that arises
due to the logistic functional form.  With scalability in mind we
adopt the Polya-Gamma latent variable method previously used for
sampling in a multinomial logistic (MNL) regression framework
\citep{polson} and extend it to the NHMM in this paper. We are motivated by the
results in \cite{polson} which showed that the Polya-Gamma latent
variable method is significantly faster than alternative sampling
schemes such as those of \cite{Holmes1} and \cite{fs}. Furthermore,
because the Polya-Gamma method uses only Gibbs sampling steps this
obviates the need for extensive parameter-tuning of the sampling
algorithm, leading to a significantly simpler implementation in
software compared to methods based on Metropolis-Hastings steps for
example. We have implemented the algorithm proposed in this paper
and made this model available in the NHMM R package on the Comprehensive R Archive Network (CRAN).

The contributions of our paper are as follows. We propose a novel hidden Markov model with inhomogeneity in both the transition and emission state-dependent distributions. This model generalizes earlier NHMMs that contained either transition or emission inhomogeneity but not both. An additional significant contribution of the paper is the development of a fully Bayesian estimation scheme for this class of models. In particular we develop a scalable Bayesian sampling scheme for the logistic transition component of the NHMM, enabling these methods to be applied to much larger data sets than in prior work. Finally we demonstrate the application of the model and the Bayesian inference algorithms to a large-scale multi-decadal precipitation data set.

Section \ref{sec:model} lays out the proposed Bayesian multivariate
NHMM. A description of the Bayesian implementation of the MCMC
algorithm and the handling of missing data, predictive simulations, and
forecasting are discussed in Section \ref{sec:mc}; further modeling considerations such as variable selection and model choice for the NHMM are included in Appendix A. Local rainfall amounts for 63 stations in and around India and the exogenous variables to be downscaled are described in Section \ref{sec:app}; specific details pertaining to the exogenous variables can be found in Appendix B. Section \ref{sec:app} also provides a brief summary of rainfall modeling.  Section
\ref{sec:results} includes the analysis and results of the NHMM when
applied to the Indian rainfall data. Our findings and general
conclusions are summarized in Section \ref{sec:conc}.

\section{Bayesian Multivariate Non-Homogeneous Markov Model}  \label{sec:model}

\subsection{The NHMM and the Likelihood} 
The observed  multivariate time-series $y_{ts},  s  = 1,\ldots,S$, with discrete-time index $t=1,\ldots,T$, is modeled using an NHMM. A general way to introduce exogenous dependence into the transition matrix is to allow each transition to have its own set of logistic coefficients (or at least $K-1$ of them, subject to identifiability), implying $O(K^2 B)$ coefficients in total for $B$ exogenous variables. This was the approach taken in \cite{meligkotsidou2011forecasting} for $K=2$.  However,  this model requires a large number of parameters as $K$ grows. A more parsimonious approach (and the one we follow in this paper) is to have one set of regression coefficients for each state, with $O(K B)$ coefficients in total,  allowing the probability of entering each state $j$ to be modulated by a set of $K$ weighted regressors (via the logistic link), but where the modulation is independent of the previous state $i$ (e.g., \cite{Kirsh04}). The intuitive interpretation is that the exogenous variables control how likely the Markov chain is to enter each state $j$ (via the logistic link and regression coefficients), and thus, as the exogenous variables change over time, so do the probabilities of being in each state. As a simple example, if one of the exogenous variables reflects seasonality (time of year), this allows the model to visit hidden states in a seasonal fashion. The hidden process ${\bm z}$  is Markov with an inhomogeneous $K \times K$ transition matrix ${\bm Q}_t$ with components $q_{ijt}$, $i,j \in \{1,\ldots,K\}$. The transition probability entries at time $t$ are modeled via a multinomial logistic link function: 
\begin{equation}\label{eq1}
q_{ijt}=P(z_{t}=j|z_{t-1}=i, {\bm x}_t, {\bm \zeta}) = \frac{\exp(\xi_{ij}+{\bm x}'_t {\bm \rho}_{j})}{\sum_{m=1}^{K} \exp(\xi_{im}+{\bm x}'_t {\bm \rho}_{m})}
\end{equation}  
where ${\bm x}_t$ is a $B$-dimensional exogenous covariate time series, $t=1,\ldots,T$ and ${\bm \rho}_{j}$ is a $B$-dimensional vector of coefficients corresponding to the $B$ components of \TR{${\bm x}_{t}=({x}_{1t},\ldots,{x}_{Bt})$}.  For notational convenience let ${\bm \zeta}=\zeta_{ij}=({\bm \rho}_j,\xi_{ij})$ for all $i,j \in \{1,\ldots,K\}$. We assign one of the ${\bm \zeta}_{\cdot j}$ to zero for some value of $j$ (one  ${\bm \rho}_j$ and a vector of ${\bm \xi}_{\cdot j}$ for some j) for identifiability. The choice of the logistic function above is discussed further in  Section \ref{zetas}. 

The other main component of an NHMM is the set of state-dependent emission distributions  $f({\bm y}_t | z_t = k,{\bm \theta}), k= 1,\ldots,K$, and where ${\bm \theta}$ is the set of all parameters of the emission distribution. Each combination of state ($k$) and station ($s$)  has its own emission distribution (for this particular application the emission distribution will be a zero-inflated mixture of exponential distributions.) In general these distributions can be specified to be inhomogeneous over time by allowing the parameters to depend on  time-varying exogenous variables  \TR{ ${\bm w}_{t,s}$,  yielding $f({\bm y}_t | z_t, {\bm \theta})$.}

\begin{figure}[htp]
\vspace{0.1in}
\begin{center}
\includegraphics[scale=0.75]{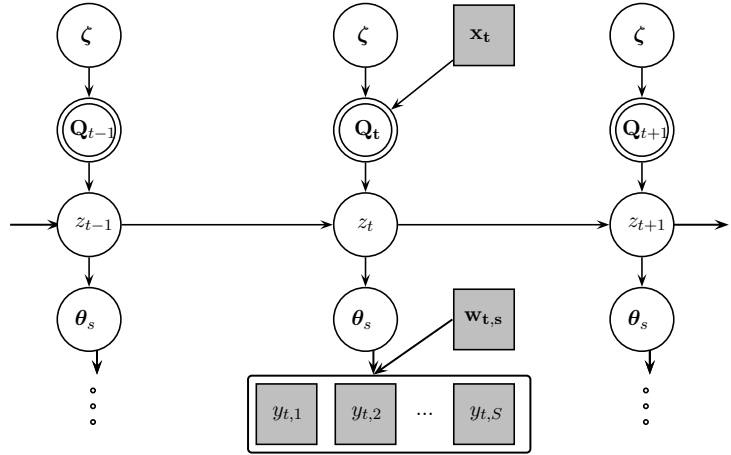}
\vspace{-6.1in} 
\end{center}
\caption{Graphical Model: The observed values ($y_{ts}, {\bf w}_{ts}, {\bf x}_{t}$) are in gray boxes. The unknown parameters (${\bm \theta}$ for the emission distribution and ${\bm \zeta}$ for the transition probabilities) and hidden states ($z_t$) are circles. ${\bf Q}_t$ (in double circles because they are directly calculated in contrast to sampled parameters) is a set of matrices that contain the transition probabilities arising from the Markov property of the hidden states and the exogenous variables ${\bf x}_{t}$.}
  \label{fig:gm}
\end{figure}

Figure \ref{fig:gm} shows a graphical representation of the multivariate NHMM and how the two types of exogenous variables (${\bm x}_t$,${\bm w}_{ts}$) impact the model. If the values of the latent variables ${\bm z}$ are assumed to be known, the conditional likelihood for the model above can be expressed as:   
\begin{equation}
P( {\bm y}_t | {\bm x}, {\bm w}, {\bm z}, {\bm \zeta},  {\bm \theta})
= \prod_{t=1}^T f({\bm y}_t | z_t , {\bm w}, {\bm \theta}) P(z_{t} |z_{t-1}, {\bm x}_t, {\bm \zeta})
\end{equation}
where  $P(z_{t} |z_{t-1}, {\bm x}_t, {\bm \zeta})$ for $t=1$ is defined via an initial state distribution $P(z_1)$ and where the $P(z_{t}|\ldots)$ transition probabilities are defined as in Equation \ref{eq1}. When the latent variables are unknown, the likelihood $P( {\bm y} | {\bm x} , {\bm w},  {\bm \zeta},  {\bm \theta})$ can be computed by marginalizing over the unknown ${\bm z}$ values in the usual recursive manner for HMMs (e.g., see \cite{scott02}). Priors and inference procedures for the unknown parameters ${\bm \zeta}$ and ${\bm \theta}$ are described in the next section.


\section{Bayesian Inference and MCMC Algorithm}  \label{sec:mc}
We describe below how to perform inference in a  Bayesian framework for the model in the preceding section using a Markov chain Monte Carlo (MCMC) algorithm. Posterior full conditional distributions can be computed for each of ${\bm z}$, ${\bm \zeta}$, and $ {\bm \theta}$ independently, such that each step of the MCMC algorithm focuses on only one set of parameters at a time. Our primary emphasis below is on the development of a sampling method for the transition matrix parameters ${\bm \zeta}$ since this has traditionally presented difficulties in the context of Bayesian analysis of NHMMs and has effectively limited the sizes of data sets that can be analyzed in past studies.   
 
If the posterior full conditional distributions are known in closed-form
then the parameters can be sampled by Gibbs steps within the MCMC. For problems
where the posterior distributions are not conjugate, it is sometimes
possible to have auxiliary variable methods facilitate rendering full
posterior conditional distributions in a form that can be sampled
from. For the NHMM described above, two sets of latent variables can
be added to the model: one set for sampling the coefficients
${\bm \zeta}$ associating with the transition probabilities of the
hidden states \citep{polson} and another set associated with
parameters ${\bm \theta}$ of the emission distributions \citep{AandC}.
Using auxiliary variables (and the resulting Gibbs sampling algorithm)
in this manner can be more efficient compared to alternative
approaches such as Metropolis-Hastings, e.g., leading in some cases to
better mixing (and thus less thinning and fewer iterations) as well as
having the advantage of not requiring tuning parameters for the
sampler (i.e., which results in a user friendly R Package).

\subsection{Sampling the ${\vect \zeta}_{k}$ Coefficients}   \label{zetas} 
In this NHMM, there are $K-1$ coefficients (${\bm \zeta}$) associated with each of the $B$ observed daily variables ${\bf x}_{t}$. These coefficient parameters (${\bm \zeta}$) are related to the transition probabilities associated with the hidden states through a link function. There are two standard link functions that are typically used in this context: the logistic multinomial (MNL) and the multinomial probit (MNP) \citep{Rii, GP-Logit}, both of which are commonly used in regression modeling of polychotomous response variables.  Although there has been relatively little literature on Bayesian inference with MNP or MNL link functions for NHMMs, Bayesian implementations  in the context of regression modeling are well-studied (e.g., see \cite{AandC,Aitch, chibgreen, RMNP, McC2, Johndrow, zhang} for MNP regression and \cite{Holmes1,Holmes2,Scott11, polson, f-s, obrien} for MNL regression).  Although mathematically quite similar \citep{paap} MNL and MNP require  quite different Bayesian sampling algorithms. The Bayesian implementation of the MNP regression model usually samples the coefficients using latent variables; this is quite efficient and therefore works well for large data sets \citep{AandC}. However, unlike the regression case, the NHMM has the additional need to calculate the transition probabilities for which there is no analytic solution in the case of the MNP. For this reason the MNL has tended to be the link function of choice for NHMM modeling. But the Bayesian implementation of MNL tends to be slow, requiring multiple tuning parameters and long sampling runs. For this reason, it is often only used with relatively small data sets and small numbers of coefficients \citep{Scott11, f-s, obrien}. For example, \cite{meligkotsidou2011forecasting} construct  a Bayesian NHMM inference procedure by drawing from the MNL regression method of \cite{Holmes1,Holmes2} using a relatively complex slice sampler to analyze a small univariate time series.
 
\cite{polson} has recently introduced a new MNL method using
Polya-Gamma latent variables, providing an algorithm that is more
efficient (both in terms of time per run and needing no tuning
parameters), which opens up the possibility of handling much larger
data sets with these models.  This provides the motivation to apply
the \cite{polson} Polya-Gamma MNL latent variable method to the NHMM.
There are a number of aspects of the MNL regression method that are
altered in the extension to the NHMM case (refer to Section 5 of
\cite{polson} for details for sampling ${\bm \zeta}$ of the MNL
regression). In the NHMM, there is no observed multinomial data as in
MNL regression. Instead, the sampled hidden states $z_t$ are set up in
matrix form to conform to the MNL regression method. ${\bf Z}$ is a
$T$ by $K$ matrix with entries $Z_{tk}$, where the columns contain
the binary representation of the hidden states (a 1 in the column of
the $z_t$ and 0 elsewhere) and is updated during
each of the iterations of the MCMC sampler. The exogenous variables
and the Markov dependence (from $z_{t-1}$ for $t=2,\ldots,T$) are
included in the matrix ${\bf X}$ which has dimension $T$ by $K+B$. The
first $K$ columns encode the information of the Markov property
($z_{t-1}$) in a binary form followed by $B$ columns for the exogenous
variables ($x_{b,t}$ for all $b$ and $t$). ${\bm \zeta}$ is a $K$ by
$K+B$ matrix of coefficients, indexed by where $k=1,\ldots,K$ and $h=1,\ldots,K+B$. One of the rows of ${\bm \zeta}$ is
set to zero for identifiability (the first $K$ by $K$ entries are the
${\bm \rho}$'s and the next B columns are the ${\bm \xi}$'s). The
  full conditional posterior distribution for the $\zeta_{kh}$'s
  allows them to be drawn conditioned on the current draw of hidden
  states and other variables. The likelihood for $\zeta_{kh}$ is given by:
\begin{align}
l({ \zeta}_{k,h}|{ \zeta}_{-k,h})&=\prod_{t=1}^T \left(\frac{e^{\eta_{tkh}}}{1+e^{\eta_{tkh}}}\right)^{Z_{tk}}\left(\frac{e^{\eta_{tkh}}}{1+e^{\eta_{tkh}}}\right)^{1-Z_{tk}}  \nonumber \\
&=\prod_{t=1}^Te^{(Z_{tk}-1/2)\eta_{tkh}}e^{-\eta^2_{tkh}/2}\omega_{tkh}PG(\omega_{tkh}|1,0)
\end{align}
where $\eta_{tkh}=X_{th}\zeta_{kh}-C_{tkh}$ with $C_{tkh}=\log\sum_{i\neq k}\exp  X_{th} \zeta_{ih}$ (which is needed for the multinomial logistic form). ${\bm \omega}$ is a set of latent variables with components $\omega_{tkh}$. At each time step there is only one observation of the hidden state, so in terms of the MNL regression the observation count is one.
The full conditional posteriors are given by: 
\begin{equation*}
     {\zeta}_{kh}|{\bm \Omega}_{kh} \sim N(m_{kh}, V_{kh}) \mbox{\hspace{.5cm}  and  \hspace{.5cm} } \omega_{tkh}|{\zeta}_{kh} \sim PG(1,\eta_{tkh})
\end{equation*}
where scalars $V_{kh} =({\bf X}'_{h}{\bm\Omega}_{kh}{\bf X}_{h}+b_{kh}^{-1})^{-1}$ and  $m_{kh} = V_{kh} ({\bf X}_h'(({\bm Z}_{k}-1/2)-{\bm \Omega}_{kh}{\bm C}_{kh})+b_{kh}^{-1}a_{kh})$. ${\bm \Omega}_{kh}$ is a T by T diagonal matrix containing $\bm\omega_{kh}$ along the diagonal. $a_{kh}$ and $b_{kh}$ are parameters of the conjugate prior; the implementation in our R package allows for a conjugate prior of the form ${\bm \zeta}_{kh} \sim N(a_{kh},b_{kh})$. If a non-informative prior is desirable then we can let $a_{kh}$ and $b_{kh}^{-1}$ be zero as we do in our rainfall example later in the paper.

The transition matrix is a necessary part of the NHMM, not typically used in MNL/MNP regression. Once the coefficients (${\bm \zeta}$) are sampled then the transition probabilities can be easily obtained through the logistic relationship given in Equation \ref{eq1}. This leads to a K by K transition matrix for time $t$:
 \begin{equation*}
{\bf Q}_t = \left[ \begin{array}{llll}
q_{11t} & q_{12t} & ... &q_{1Kt}\\
q_{21t} & q_{22t} & ... &q_{2Kt}\\   
        &         &     &         \\
q_{K1t} & q_{K2t} & ... &q_{KKt}\end{array} \right]
\end{equation*}  
where each row of ${\bf Q}_t$ sums to one.

\subsection{Sampling the Hidden States, Conditioned on Parameters ${\vect \zeta}$ and ${\vect \theta}$}

Conditioned on sampled values of the parameters ${\bm \zeta}$ and ${\bm \theta}$, and given the observed data ${\bm y}, {\bm x},$ and ${\bm w}$, the posterior full conditional distribution of the hidden state $z_t^n$ at the $n$th sampling iteration is as follows (dropping the third subscript $t$ from the $q$ variables for clarity):
\begin{align*}
z_{t}^n|{\bm \zeta},{\bm \theta},\dots &\sim 
Multi\left(  
\frac{q_{z_{t-1}^n,1} \ \ q_{1,z_{t+1}^{n-1}} \ \  f_{1}(.)}
{   \displaystyle\sum_{k=1}^{K} q_{z_{t-1}^n,k} \ q_{k,z_{t+1}^{n-1}} \ f_{k}(.)},\dots,\frac{q_{z_{t-1}^n,K}\ \ q_{K,z_{t+1}^{n-1}} \  f_{K}(.)}
{\displaystyle\sum_{k=1}^{K} q_{z_{t-1}^n,k}\ q_{k,z_{t+1}^{n-1}} \ f_{k}(.)} 
\right)
\end{align*}
where $f_{z_t}(.)=f({\bm y}_t|z_t = k, {\bm \theta})$ is the emission distribution for state $k = 1, \ldots, K$. Each of the $z_t$ are sampled in succession for all $t=2,\ldots,T$ at each of the $n=1,\ldots,N$ iterations of the larger MCMC algorithm. Without loss of generality, we assign the first hidden state, associated with day one of the time series, to state one: $Pr(z_{t=1}=1)=1$.

We can sample the hidden states ${\bm z}$ using well-known efficient recursive techniques. For example, \cite{scott02} describes two Bayesian algorithms for sampling the hidden state of an HMM: a forward-backward (FB) recursive algorithm and a direct Gibbs (DG) sampler. The FB method mixes more rapidly but takes more computational effort. We use the DG method, which can require more iterations (for better mixing) but is less expensive per iteration.

Finite mixture models, including NHMMs, can suffer from the issue of non-identifiability of the hidden states \citep{jasra05, spezia09}. Any pair of states could swap labels and the likelihood would remain invariant; leading to identical marginal posterior densities, see \cite{fruhwirth2006finite} for a full discussion. Both \cite{scott02} and \cite{meligkotsidou2011forecasting} discuss this issue for similar HMM and NHMM models respectively. However, NHMMs are less likely to suffer from label switching compared to HMMs or finite mixtures due to the dependence of the latent states on fixed covariates, which effectively makes label-switching less likely for the states. In particular, for the model we propose in this paper, both the state transitions and the emission distribution parameters are dependent on fixed covariate time-series. In our experimental results with rainfall data (described in Section \ref{sec:results}) we did not see any evidence of label switching. 


\subsection{Sampling for the Emission Distribution Parameters}
For this application we model daily rainfall amounts by a zero-inflated mixture of two exponential distributions, an approach that has been found most effective in past work \citep{wool, wilksc, wilksa, wilksb,ailliotrev}. Other possible modeling options include zero-inflated Gamma distributions or mixtures of exponential, Normal, or Poisson distributions \citep{hay91, hughes94, charles99, bellone00, holsclaw2015bayesian}. The zero-inflated mixture of two exponential distributions has a physical interpretation of its three components corresponding to no rain, light rain, and heavy rain. The delta function at zero ($\delta_0$) allows for zero inflation for additional dry days, and the light rain and heavy rain each have an exponential distribution, where:
\begin{equation}
y_{t,s}| z_t,{\bm \theta} \sim
  p_{0ts}\delta_0+p_{1ts}Exp(\lambda_{1z_ts})+p_{2ts}Exp(\lambda_{2z_ts})
  \end{equation}
  where $z_t=k$ and for this application ${\bm \theta}$ denotes the mixing       probability parameters and rate parameters of the emission distributions. The mixing probabilities ${\bf p}=(p_{0ts}, p_{1ts},p_{2ts})$ are assumed to be dependent on the $A$ exogenous variables ${\bm w}_{t,s}=(w_{1ts},\ldots,w_{Ats})$ and are modeled by a generalized linear model (GLM) through a probit link: ${\bf p}_{t,s}=g^{-1}(\beta_{0z_ts}+{\bm w}'_{\cdot ts}{\bm \beta}_{1 \cdot s})$ for all $a$; $\beta_{0z_ts}$ provides the dependence on the $K$ hidden states, with $z_t=k$ and $w_{ats}\beta_{1as}$ as the mean. Let ${\bm \beta}=(\beta_{0z_ts},\beta_{1as})$ for $t \in T$, $a \in A$, and $s \in S$; let ${\bm \theta}=({\bm \lambda}, {\bm \beta})$ denote  the parameters of the emission distributions. The $\beta_{0z_ts}$ are state dependent and function like a random effect whereas the $\beta_{1as}$ are not state dependent thus allowing significance testing of the exogenous variable per station.

The probit link for ordered multinomial categories allows for the sampling of the coefficients to be done through the standard Bayesian data augmentation approach
\citep{cox71,mccullagh1989, AandC}. To allow for conjugate full
conditional posterior distributions of the parameters of
${\bm \beta}$, we need to introduce two sets of latent variables
(${\bm L}$ and ${\bm M}$). The first set of latent variables ${\bm L}$
(with components $L_{ts}$ taking values in the set $\{0,1,2\}$)
facilitates calculations of ${\bf p}$. The emission distribution
becomes: 
\begin{align*}
y_{ts}|...&\sim p_{0ts} \delta_0+p_{1ts} Exp(\lambda_{1z_ts})+p_{2ts}Exp(\lambda_{2z_ts})\\ 
&\sim [\delta_0I_{L_{ts}=0}][Exp(\lambda_{1z_ts})I_{L_{ts}=1}][Exp(\lambda_{2z_ts})I_{L_{ts}=2}]
\end{align*}
where $z_t=k$.  A second set of latent variables ${\bm M}$ with components $M_{ts} \sim N(\beta_{0z_ts}+{\bm w}'_{\cdot ts}{\bm \beta}_{1 \cdot s},1)$ is introduced to enable Gibbs sampling of ${\bm \beta}$. The latent variables ${\bm L}$ are three ordered categories (no rain, light rain, and heavy rain) which, following the ordered multinomial probit algorithm in \cite{AandC}, requires one fixed break point (set to zero) and one unknown break point ($\gamma$) (more categories would require more unknown breakpoints). The relationship between ${\bm L}$ and ${\bm M}$ is as follows:
\begin{equation*}\label{vtoM}
L_{ts}=
\begin{cases}
          0 &   M_{ts}<0             \\
          1 &  0 < M_{ts} < \gamma \\
          2 &  \gamma < M_{ts}   
\end{cases}
\end{equation*}
 This results in posterior full conditional distributions as described in \cite{holsclaw2015bayesian}. For our rainfall modeling application the $\lambda_{1ks}$ and  $\lambda_{2ks}$ parameters are each given a low weight conjugate prior ($\Gamma(1,1)$) (label switching does not occur because of the ordered nature of the latent variable method of \cite{AandC}.) The ${\bm \beta}$ coefficients have non-informative priors as well \citep{AandC}. This setup leads to the parameters ${\bm \theta}=({\bm \lambda}, {\bm \beta})$ having closed form full conditional posterior distributions that can be sampled via Gibbs steps in the MCMC algorithm.

\subsection{Missing Data Imputation}\label{miss}
	The missing data points can be
  treated as unknown random variables whose posterior distributions are
  inferred along with the other variables in the model.  The posterior
  conditional distribution of each missing data point ($y^o_{ts}$ at
  time $t$ and station $s$) is given by:
  $\bm{y}^o_{t}| z^o_t,\ldots \sim f({\bm \theta}^o_{z^o_t},{\bm
    w}_{t})$. Data that is missing at random from the observed time series can be imputed
  as part of the MCMC algorithm. $y^o_{ts}$ can be drawn at each iteration of the
  MCMC from this distribution, where ${\bm \theta}^o_{z^o_t}$ and $z^o_t$ are also draws from their posterior full conditional distributions.


\subsection{Predictive Conditional Chains and Forecasting}\label{pred}
New time-series of length $T$ can be simulated conditioned on the ${\bf x}$ and ${\bf w}$ inputs. In this paper we simulate these forecast chains conditioned on the
exogenous variables for held out years of inputs. First, the exogenous variables {\bf x} and the sampled coefficients (${\bm \zeta}^o$) are used to generate the transition
probabilities (${\bf q}^*$) and then chains of the hidden states
(${\bf z}^*$) are simulated. Unlike the scheme for imputing missing data described in Section \ref{miss}, the predictive conditional chains require a predictive draw from the
hidden states (${\bf z}^*$). Because of the autoregressive nature of
the states (the Markov property of the NHMM), the conditional
predictive chains can be generated one day at a time, dependent on the
previous day. The exogenous variables {\bf w}, their
sampled coefficients (${\bm \theta}^o$), and the newly generated
chains of hidden states are then used to simulate from the emission
distribution (${\bf y}^*_r$). For a new time step $r$, this process
can be expressed as:
\begin{align*}
   q^*_{ijr} | {\bf X}_r, {\bm \zeta}^o &= g^{-1}({\bf X}'_r{\bm \zeta}^o)\\
	 z^*_r| q^*_{z^*_{r-1}jr}  &\sim Multi(q^*_{z^*_{r-1}1r},\ldots, q^*_{z^*_{r-1}Kr})\\
	 {\bm y}^*_r|  z_r,\ldots &\sim f_k({\bm \theta}^o_{z^*_r},{\bf w}_r)
	\end{align*}
	where $z^*_r=k$ and $q^*_{ijr}$, $z^*_r$, and ${\bm y}^*_r$ are new predictive draws at time $r$.

Specifically, we use the first 27 years of data (1981-2007) to fit the model and then generate predictive conditional chains for  2008-2010. These chains can then be compared to three years of held out observed ${\bm y}$ data for the  purposes of model selection and distributional checks (see Section \ref{mc} and also \cite{supplementA} for plots.)

\section{Analysis of Daily Rainfall in India}\label{sec:app}

India has a large population that relies heavily on annual monsoonal rainfall patterns. Variations in rainfall occurrence and amounts can lead to  floods and droughts with significant major impacts on food production, hydroelectricity production, and human safety.  These variations can be better understood by studying the interactions  of daily rainfall with large scale and regional exogenous weather variables \citep{wilks99,ImmerzeelEtAl_Science10, HansenEtAl_CR06}. The daily time scale for rainfall modeling is of particular interest because of the effect of flooding, dry spell length, and soil moisture content on agriculture and food supply \citep{sterncoe}.

\subsection{Rainfall Data}
The rainfall data used in this paper (as briefly described earlier in Section 1) corresponds to daily rainfall amounts\footnote{Data obtained from the U.S. National Centers for Environmental Prediction (NCEP) Climate Prediction Center (CPC) Global Summary of the Day (GSOD) Observations.} between the years of 1981-2010 for   a diverse set of 63 weather stations in the Indian region (Figure \ref{statloc}).  Stations were selected for inclusion in the data set if no more than 10\% of the days for that station had missing observations. This resulted in a total of 689,850 observations over the 30-year period (with leap days removed as in \cite{furrer07}), with 63 daily rainfall time-series ${\bm y}_{ts}, 1 \le s \le 63, 1 \le t \le 10950$.

\begin{figure}[htp]
\begin{center}
\includegraphics[scale=0.30]{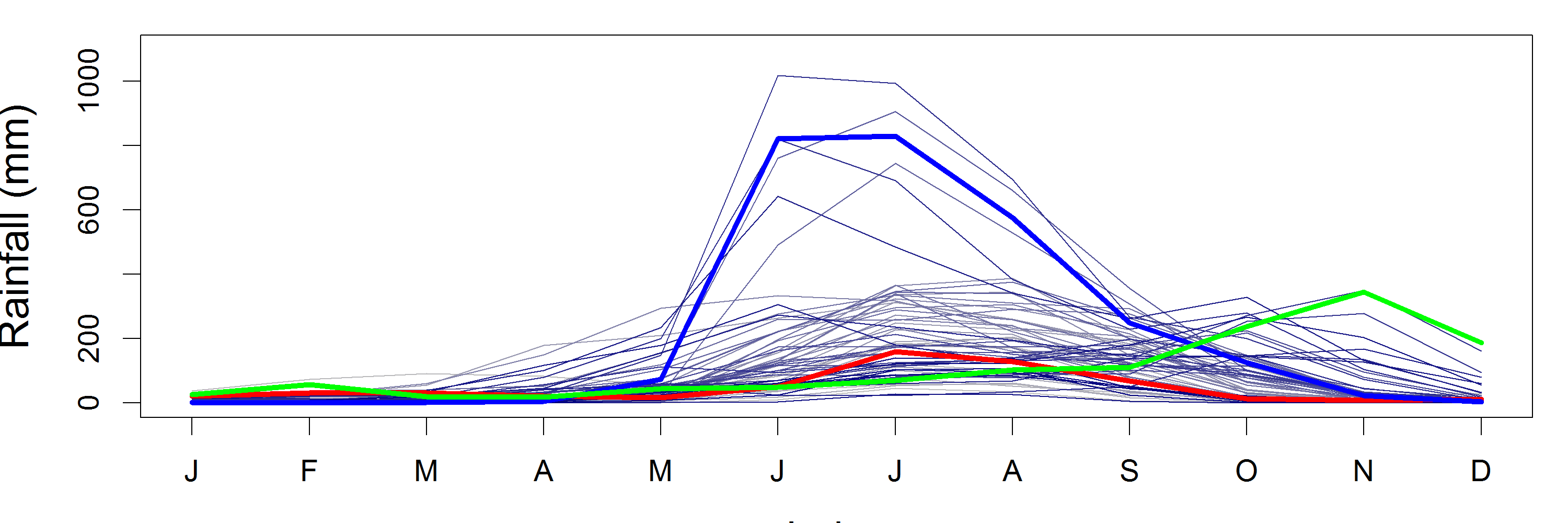}
\end{center}
\caption{Monthly rainfall (mm) averaged over all years, one line for each of the 63 stations.  Stations 3, 40, and 52 are highlighted with bold lines. See also Figures 1 and 2 for context.}
  \label{fig16}
\end{figure}

Figure \ref{fig16} shows a plot of the seasonal cycle, where each line represents one of the 63 stations, illustrating the diversity of rainfall and its seasonality across the stations. Some stations have strong summer monsoonal maxima, while others are much dryer, and some peak towards the end of the calendar year. 

\subsection{Covariate Climate Indices}
\begin{figure}[htp]
\begin{center}
\includegraphics[scale=0.55]{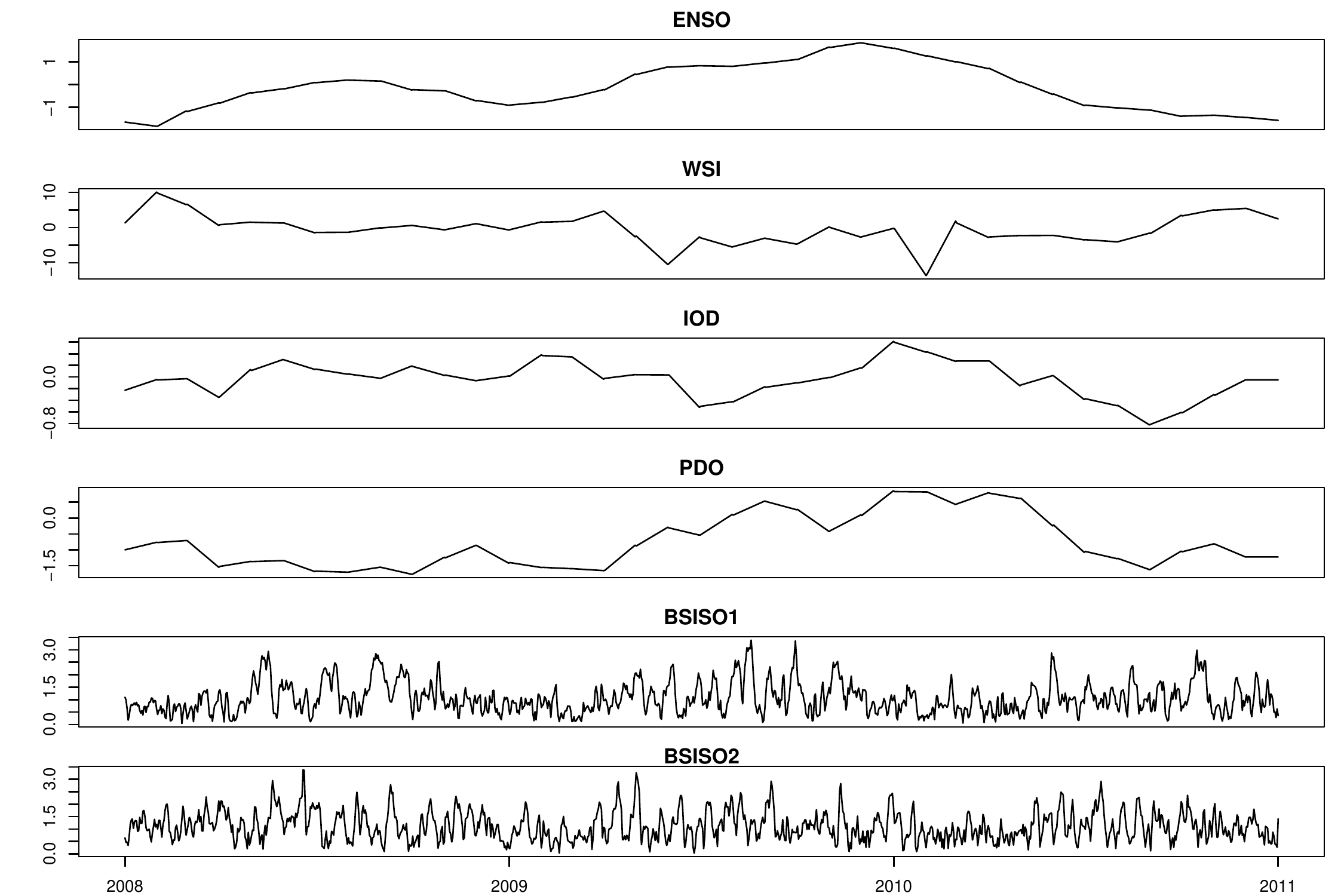}
\end{center}
\caption{Exogenous variables: ENSO, WSI, IOD, PDO, BSISO1, and BSISO2 for three years. ENSO, WSI, IOD, and PDO are calculated by linearly interpolating  monthly values to the daily time-scale. BSISO1 and BSISO2 are available on a daily basis.}
  \label{input}
\end{figure}

  The roles that  remote climate ``drivers'' play in Indian
  rainfall variability are not fully understood, especially at
  regional scales, and the potential for prediction remains a topic of active
  research \citep[e.g.,][]{MoronRobertsonGhil_CD12}.  In this context we chose six established
  climate indices for our model as exogenous variables. All
  have been shown in previous studies to be associated with
  rainfall variability over India on different time scales. The variables are Westerly wind Shear Index
  (WSI), El Ni\~no/Southern Oscillation (ENSO), Indian Ocean Dipole
  (IOD), Pacific Decadal Oscillation (PDO), and two components of the
  boreal summer intraseasonal oscillation (BSISO1 and BSISO2). The
  WSI encodes year-to-year (interannual)
  changes in the strength of the summer monsoon winds which are
  closely related with interannual variations in the monsoon
  rainfall \cite{wang99,greene11}. ENSO and IOD are known influences on
  rainfall on interannual time scales \citep{gadgil03a}, whereas
  PDO has a less well understood impact
  \citep{JosephEtAl_AOSL13}. The monsoon tends to be stronger during
  the La Nina phase, when this ENSO index is \emph{negative}
  \citep{gadgil03a} and when IOD is positive \citep{gadgil03a}. These
  aforementioned three variables are closely related to monthly SST. On sub-seasonal time scales Indian
  monsoon rainfall is impacted by the boreal summer intraseasonal
  oscillation (BSISO) for which we use the two indices BSISO 1 and 2
  defined by \cite{LeeEtAl_CD13}. Figure
  \ref{input} shows the six input time series, for the years
  2008--2010. For a more detailed explanation of each of these
  variables see Appendix B.

Understanding of these exogenous variables has been hindered by longer time scale non-stationarity, possibly associated with anthropogenic climate change, or the remote impacts of other ocean basins \citep{GershunovEtAl_JCL01}. Our approach is thus to include all six indices as candidate covariates, where BSISO is given as daily values and the monthly series (ENSO, WSI, IOD, PDO) are interpolated linearly to daily values.

In our model, there are two ways exogenous variables can be included: a station-level $A$-dimensional time series ${\bm w}$ with components $w_{ats}$ or a global $B$-dimensional time-series ${\bm x}$ with components $x_{bt}$. The station-dependent variables (${\bm w}$) are {\it local} in nature and directly influence the mixing weights of the point mass at zero and mixture of exponential distributions of the emission distribution for each station individually. Lower frequency climatic drivers tend to impact the climatic background, and we thus introduce the impacts of the WSI, ENSO, IOD, and PDO climate drivers via {\bf w}, directly influencing the characteristics of the emission distributions. In addition, a station-specific seasonal cycle (annual and bi-annual harmonic terms - four total terms) and a long term drift term are included in ${\bf w}$.

In contrast, the large scale time-dependent exogenous variables (${\bm x}$) are not station-specific---they affect the whole region and influence the transition probabilities of the hidden states of the NHMM. Indian monsoon rainfall is mostly generated by local scale thunderstorm activity and monsoon depressions, while mid-latitude western disturbances are important over northern India, especially in winter. On sub-seasonal time scales the paths and intensities of these phenomena are controlled by large scale atmospheric circulation patterns that can be naturally represented by a discrete set of weather states and the transitions between them \citep{GhilRob02}. These are modulated by the BSISO whose impacts are thus encoded in the model via the ${\bf x}$ variable influencing the transition matrix. Additionally, a general seasonal cycle is included for the state transitions in ${\bf x}$, which also has terms to fit annual and bi-annual harmonics due to seasonal cycles (i.e., there are a total of four sine and cosine terms in {\bf x}). 

\section{Results}\label{sec:results}

In this section we assess the model's ability to  capture distributional, temporal, and other aspects of rainfall, as well as investigating the effects of the exogenous climate variables through information gained from the parameter uncertainty estimates.
After fitting the model using 27 years of daily data (1981-2007) we simulated 1000 chains of length 27 years for the 63-station network, conditioned on the corresponding 27 years of exogenous variables ${\bm w}$ and ${\bm x}$, to produce the Figures in this section. The last 3 years (2008-2010) of observed data (${\bm w}$, ${\bm x}$, and ${\bm y}$) were held out. These 3 years of held out data were used for model selection (Appendix A) and also to compare with predictive conditional chains (with plots shown in \cite{supplementA}). For model selection, we use a combination of the Bayesian Information Criteria (BIC) and predictive log-probability scores (PLS) for selecting the number of hidden states and selecting among different combinations of exogenous variables. All of the results in the remainder of the paper are for the selected model with $K=7$ states which was used to generate 1000 simulated chains of 27 years of data.

\subsection{Hidden States}
\begin{figure}[htp]
\begin{center}
\includegraphics[scale=0.50]{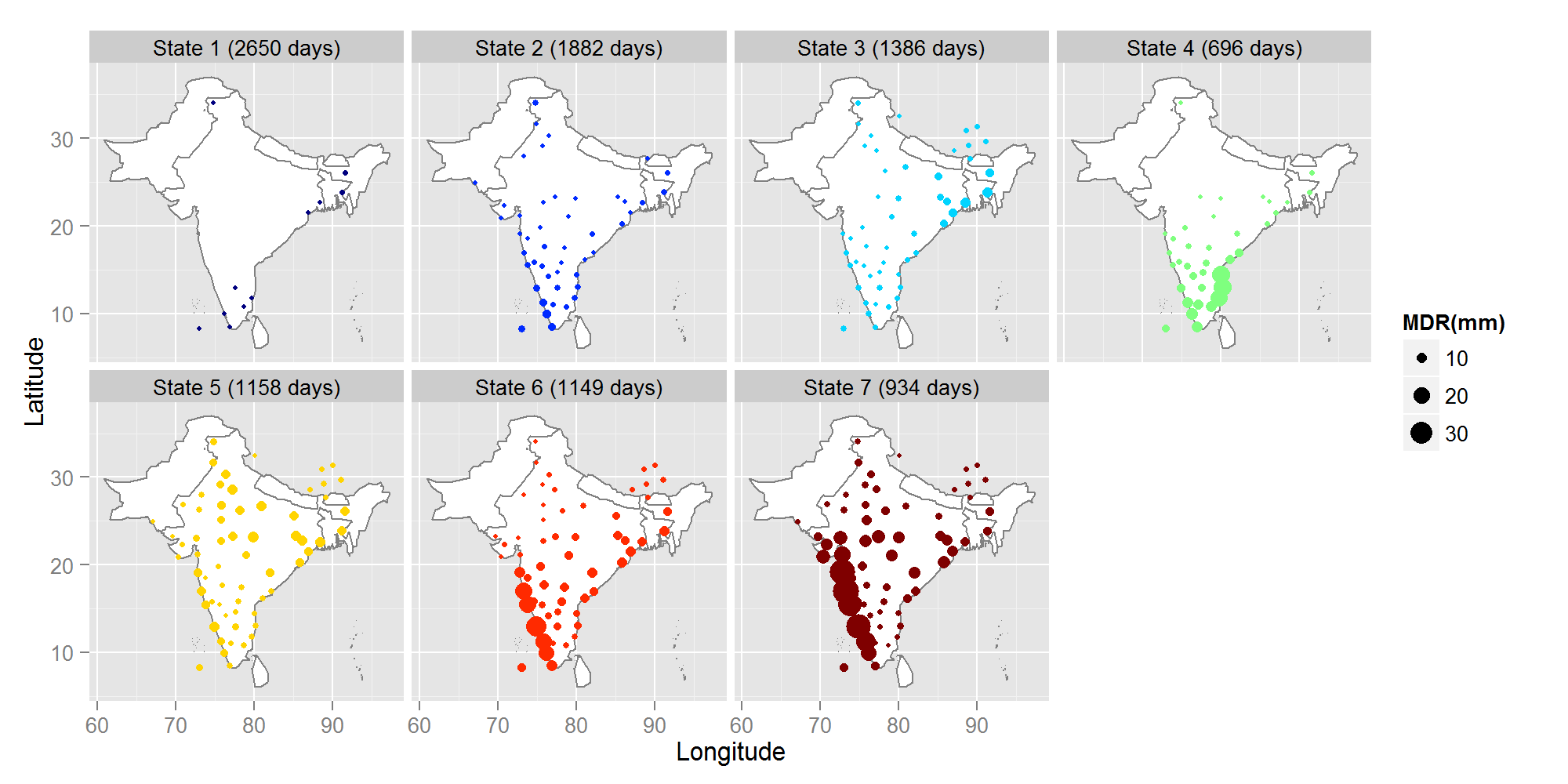}  
\end{center}
\caption{Mean daily rainfall amount for each of the hidden states,
  given by circle size.}
  \label{rain-MDR}
\end{figure}

From the MCMC algorithm, we sample the hidden states (${\bm z}$) and
marginalize over the iterations of the algorithm to find the most
probable hidden state for each day (similar to a Viterbi sequence
\citep{forney78}). Figure \ref{rain-MDR} shows the mean daily rainfall
amount at each station for each of the hidden states. The top of each
pane indicates the number of days assigned to each state given the
most probable state sequence (there are a total of
$27 \times 365= 9855$ days). State 1 represents largely dry days
across the whole domain (some stations have little to no rainfall and have no dot), with moderate rainfall occurring in states 2
and 3. State 5 characterizes heavier rainfall over north-central
India. The heaviest rainfall occurs along the western coasts in states
6 and 7, while state 4 is unique in representing rainfall over the
southeast coast. 

\begin{figure}[htp]
\begin{center}
\includegraphics[scale=0.35]{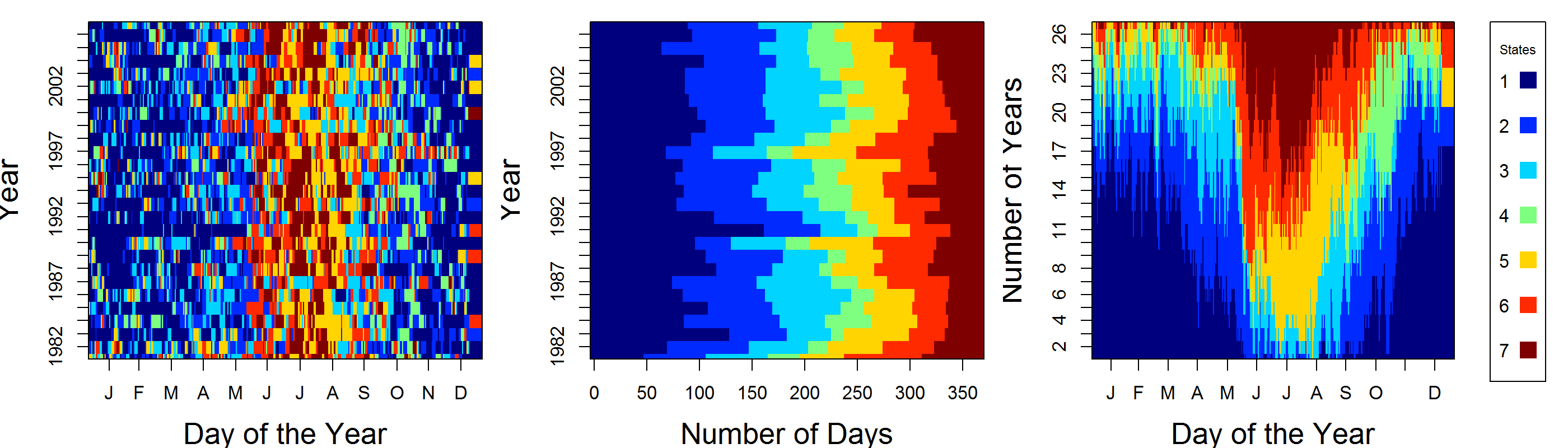}
\end{center}
\caption{Most probable sequence of hidden states (left), together with the annual averages of number of days per state (middle), and averages across years for each calendar day (right). }
  \label{z-temp}
\end{figure}

Figure \ref{z-temp} shows attributes of the daily sequences of states. The left panel shows
year-long chains of most probable state sequences for each year of the data set, illustrating the dramatic seasonality of the summer monsoon together with a large amount of sub-seasonal and interannual variability with a stochastic character. The middle panel sums along the rows to depict variability in the annual counts of each state. The right panel sums by column to depict the seasonality and has the counts of each state per day of the year given that we observed 27 years; January 1st is on the left and December 31st is on the right. The state occurrence frequencies can be seen to follow distinct seasonal patterns.  

The temporal distributions of each state can be naturally understood in terms of the rainfall climatology of India by referring to their temporal evolutions, shown in Figure \ref{z-temp}. The wetter states occur more frequently during the summer monsoon season, while state 4 is characteristic of the winter monsoon over SE India, peaking in boreal autumn (right pane of Figure \ref{z-temp}).

\subsection{Rainfall Simulations}
In this section, we show both the seasonality and distribution of rainfall for the average across all stations as well as for three specific and diverse stations. While it is straightforward to capture the seasonality and rainfall distribution at a single station, doing it jointly across multiple stations is non-trivial \citep{charles99,bellone00}. The NHMM approach provides a useful mechanism for addressing this joint modeling problem by conditioning the stations on common shared state variables. The Figures below are for the chains simulated on the first 27 years of data---see \cite{supplementA} for similar plots for the 3 held-out years of data\footnote{\cite{supplementA} also includes distributional plots showing the NHMM's ability to capture dry spell and wet spell lengths, inter-annual variability of mean rainfall, dry days, and heavy rainfall events.}.  

\subsubsection{Seasonality}
\begin{figure}[htp]
\begin{center}
\subfigure[Rainfall]{\label{rs1}\includegraphics[scale=0.30]{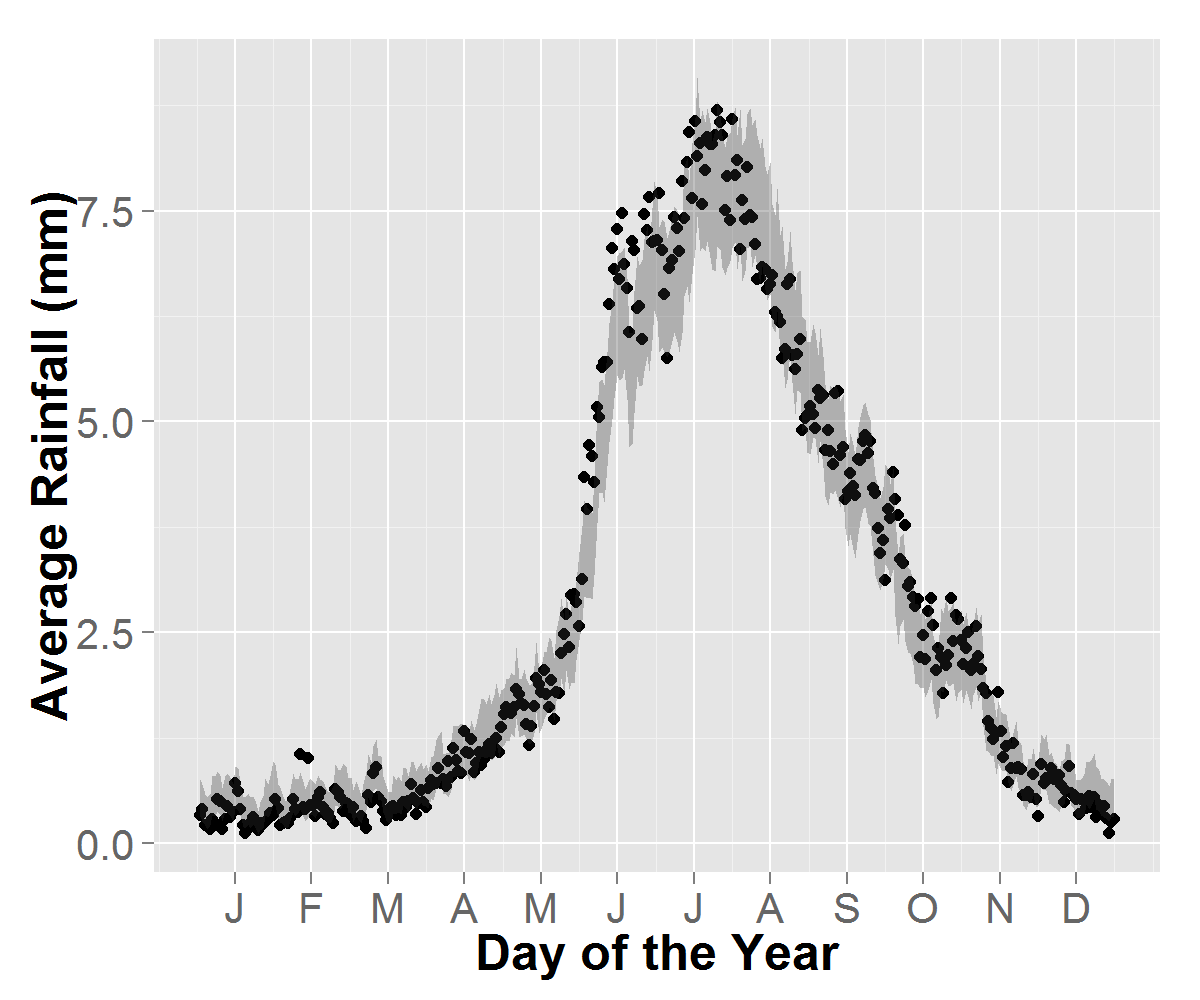}}
\subfigure[Station 3: Rainfall]{\label{rs2}\includegraphics[scale=0.30]{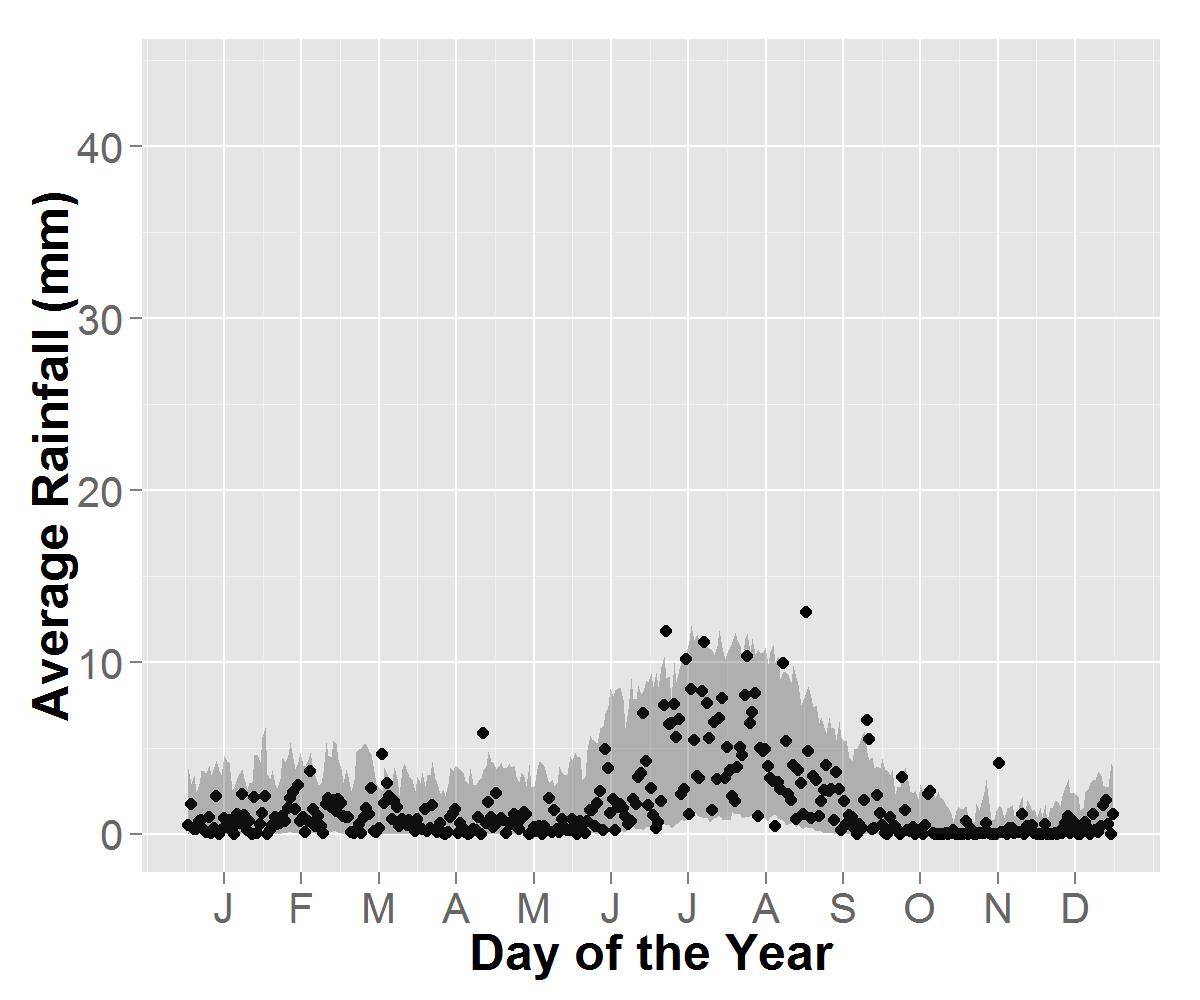}}\\
\subfigure[Station 40: Rainfall]{\label{rs3}\includegraphics[scale=0.30]{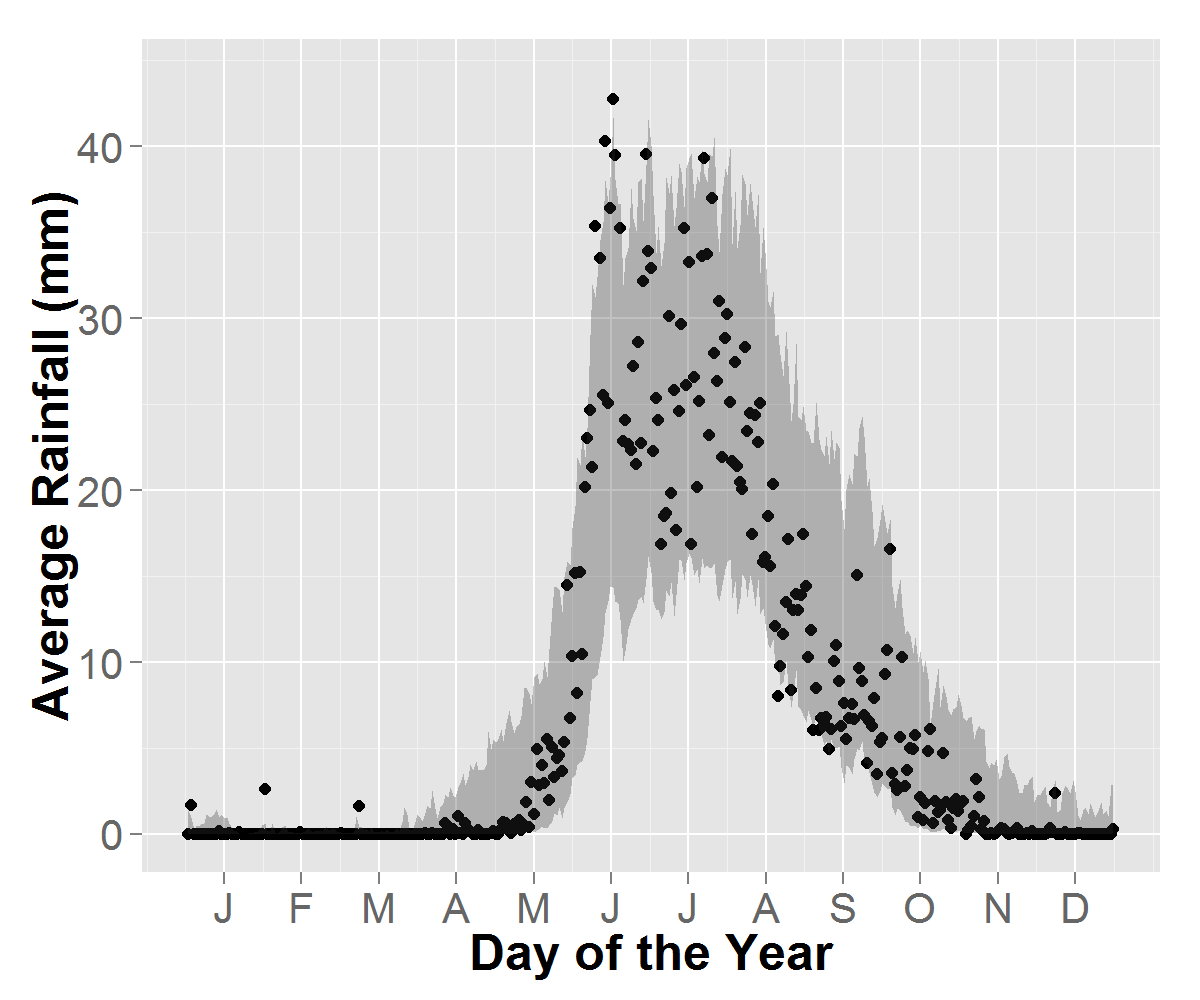}}
\subfigure[Station 52: Rainfall]{\label{rs4}\includegraphics[scale=0.30]{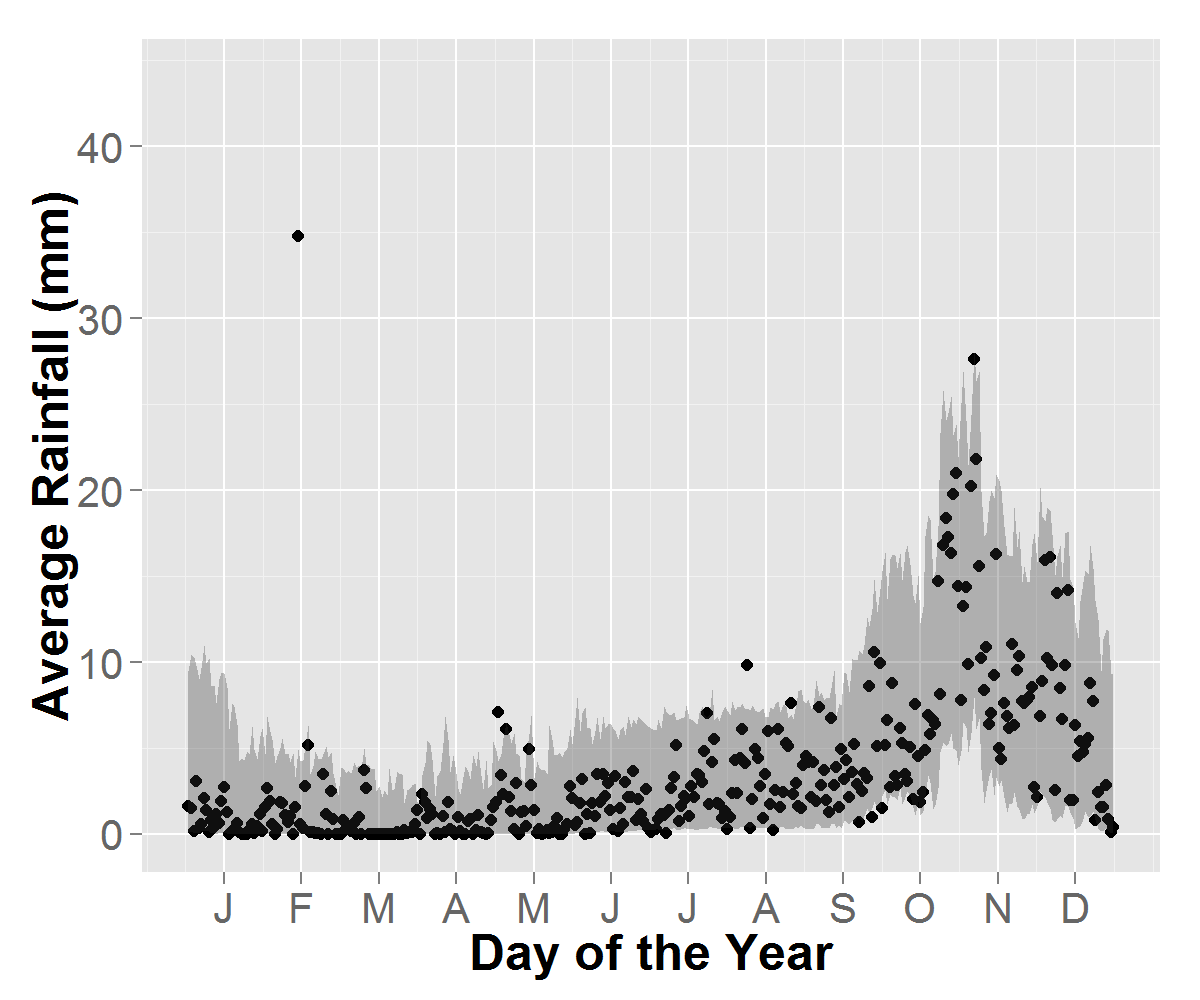}}
\end{center}
\caption{Observed data averaged over 27 years (black) and 1000 simulated sets and their 95\% PI bands (gray).  a) rainfall averaged over all stations,   b) station 3 (dry), c) station 40 (wet summer), and d) station 52 (wet winter).}
  \label{rseas2}
\end{figure}

Figure \ref{rseas2} provides an indication of how well the model captures seasonality. The observed average rainfall per day over 27 years is shown by the black points in the figure and   the NHMM simulations correspond to the 95\% probability interval (PI) bands in gray for the 1000 simulated sets. Figure \ref{rs1} shows that the seasonality of the simulated data from the model is similar to the observed data averaged over all stations. Figures \ref{rs2}, \ref{rs3}, and \ref{rs4} show the same seasonal plots for rainfall but for the three contrasting stations 3, 40 and 52. The seasonality of the simulated data is similar to the observed data, when averaged over all stations, as well as for the three individual stations with diverse climatologies. (\cite{supplementA} includes similar figures for all of the individual stations.)

\subsubsection{Distributional Checks}
\begin{figure}[htp]
\begin{center}
\subfigure[All Stations Rainfall]{\label{rld1}\includegraphics[scale=0.30]{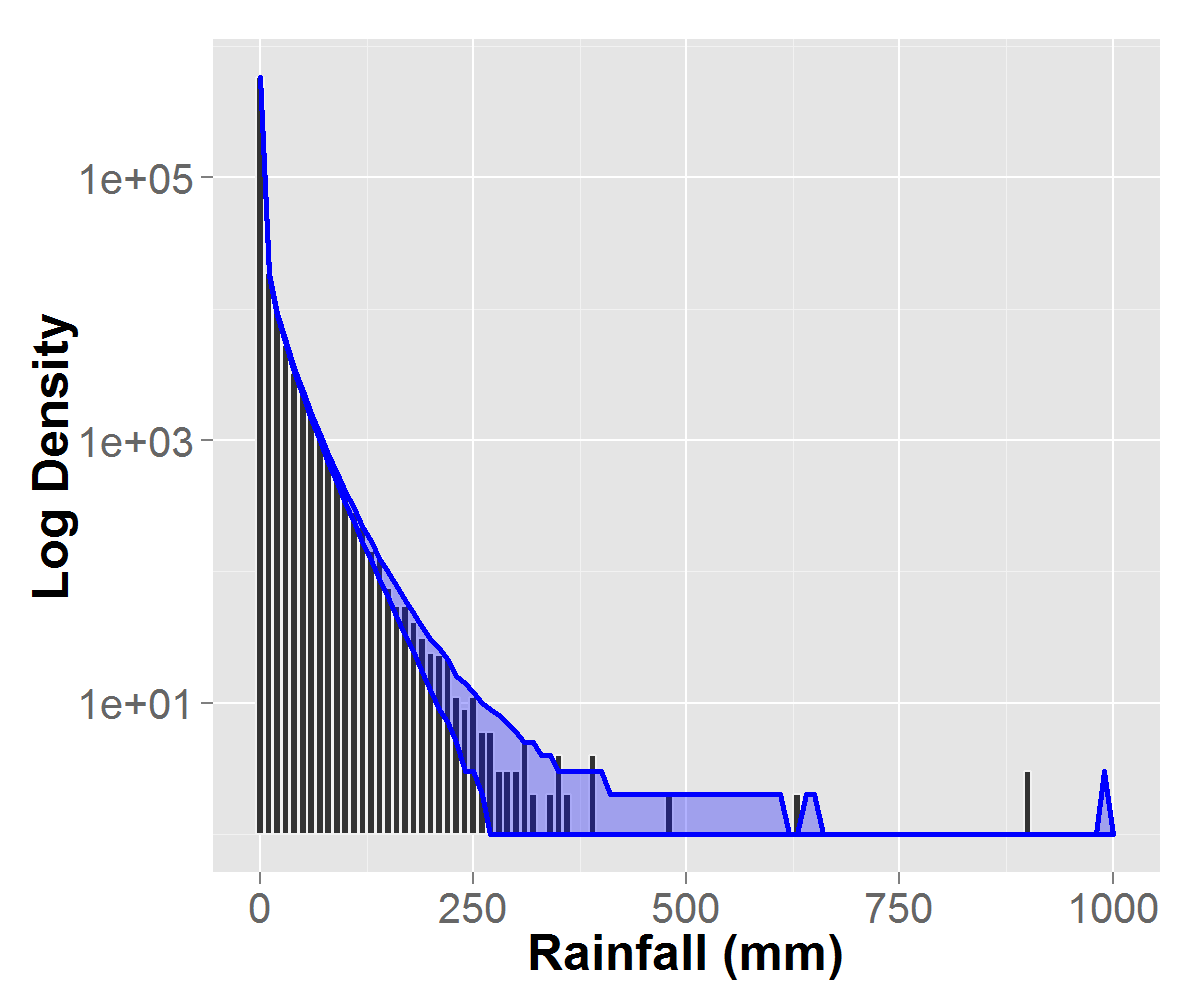}}
\subfigure[Station 3 Rainfall]{\label{rld2}\includegraphics[scale=0.30]{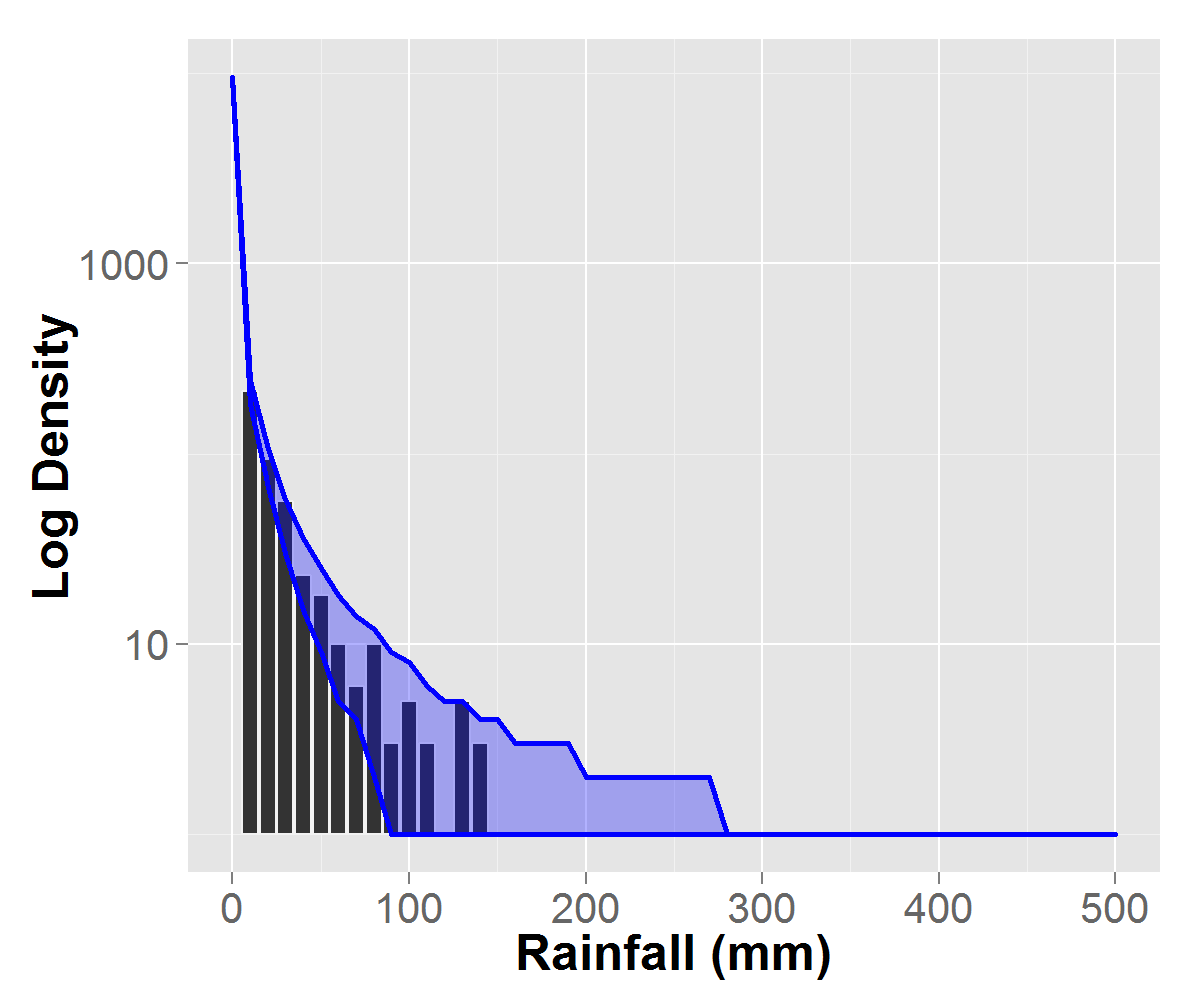}}\\
\subfigure[Station 40 Rainfall]{\label{rld3}\includegraphics[scale=0.30]{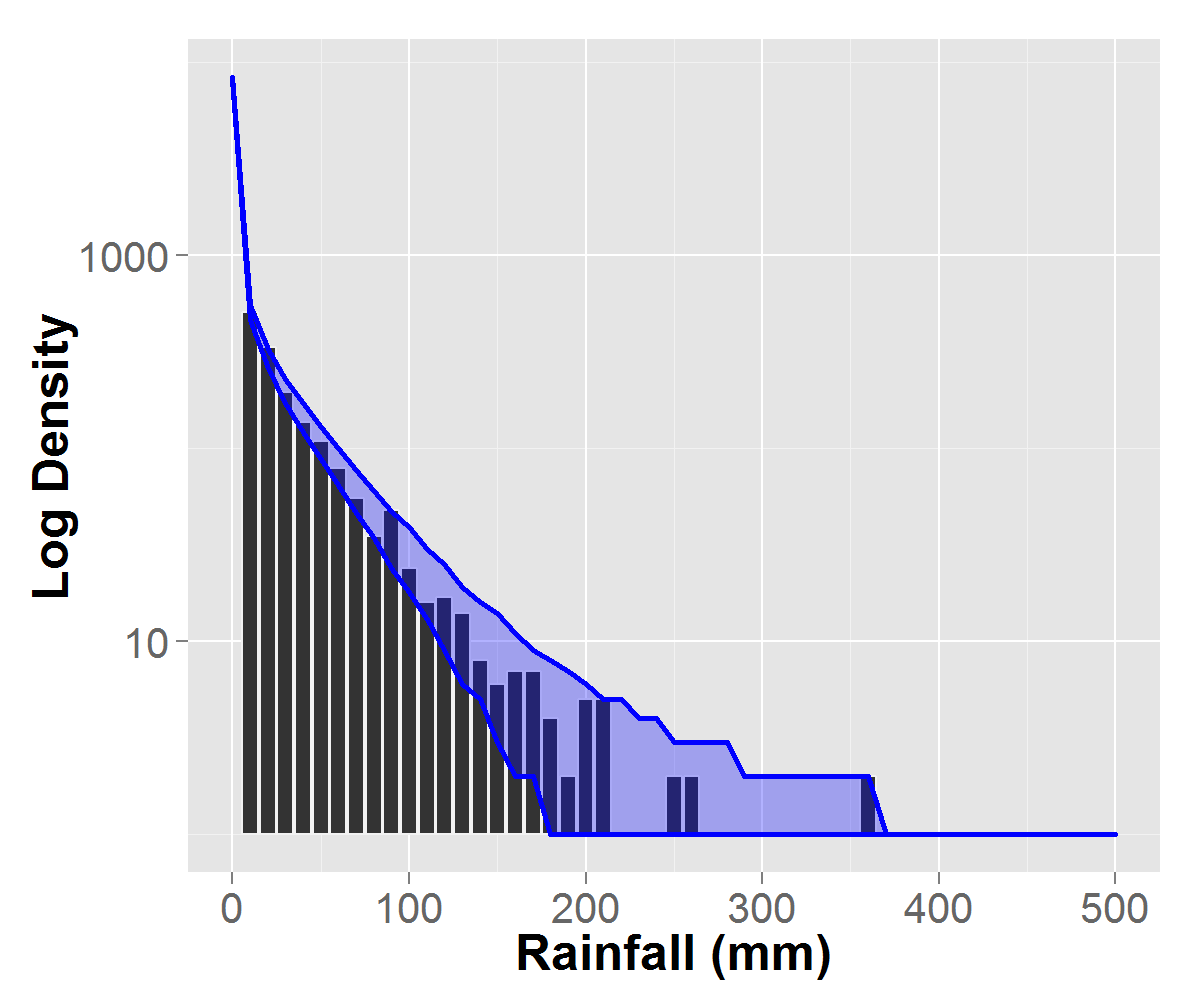}}
\subfigure[Station 52 Rainfall]{\label{rld4}\includegraphics[scale=0.30]{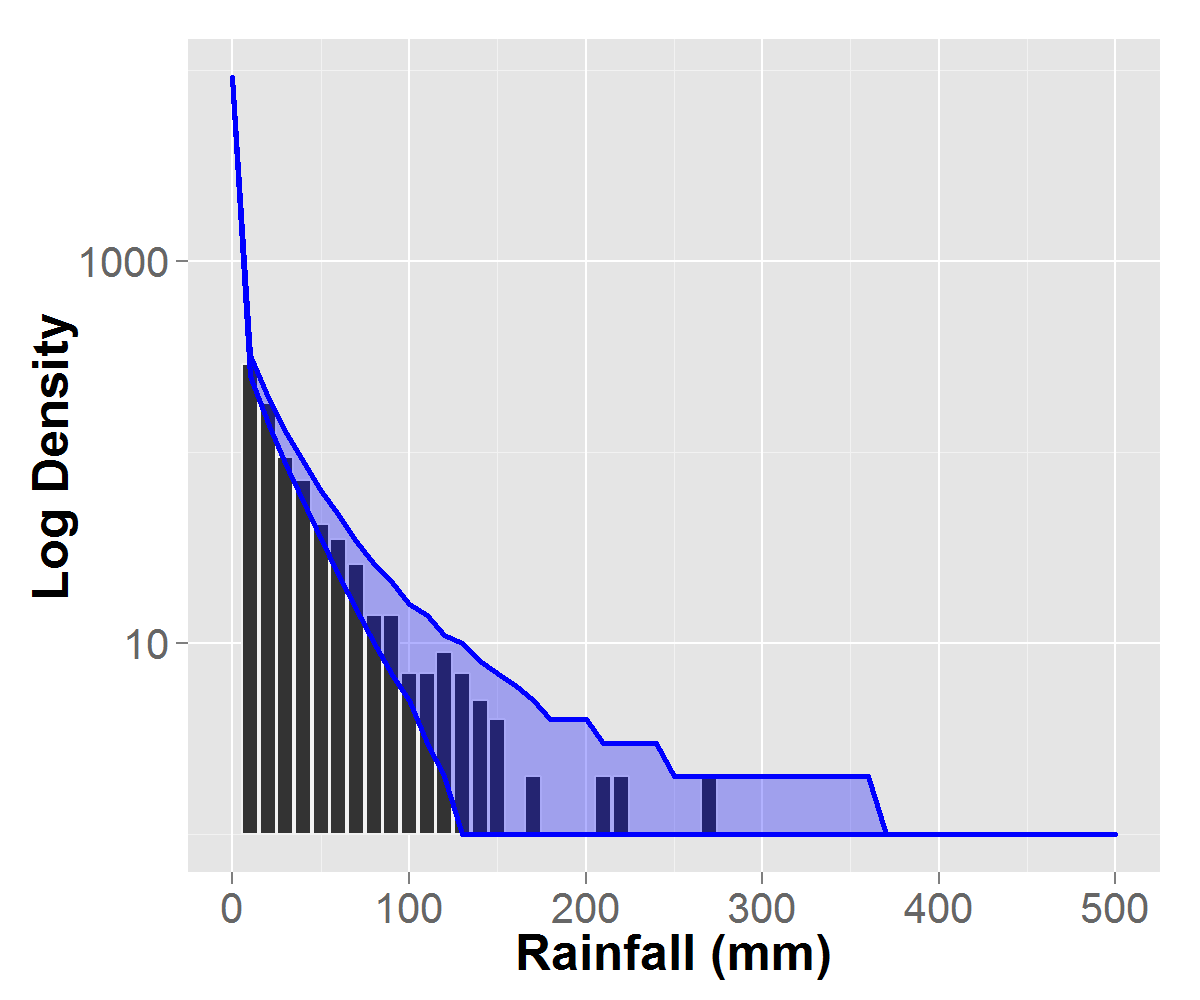}}
\end{center}
\caption{Observed data density on the log scale (gray) and 1000 simulated chains for 27 years given by the 95\% PI bands. a) rainfall averaged over all stations, b) station 3 which is drier, c) station 40 which is wetter in summer, and d) station 52 which is wetter in winter. }
  \label{tdist2}
\end{figure}

Also of interest is the NHMM's ability in capturing the large scale distributional properties of the observed rainfall data.  Figure \ref{tdist2} shows the observed 27 years of data in the gray histogram and the 95\% PI bands for the simulated chains (densities are plotted on a logarithmic scale.) Figure \ref{rld1} is averaged over all stations, Figure \ref{rld2}, \ref{rld3}, and \ref{rld4} shows the data density for the same dry and wet stations as in Figure \ref{rseas2}. These plots show that the distribution of the observed data is reasonably well represented by the model (see \cite{supplementA} for results pertaining to the dryspells for more details on the dry days distribution.) Each of the diverse stations are well modeled, from wetter to drier locations.

\subsection{Model Diagnostics of Climate Controls}
\begin{figure}[htp]
\begin{center}
\includegraphics[scale=0.40]{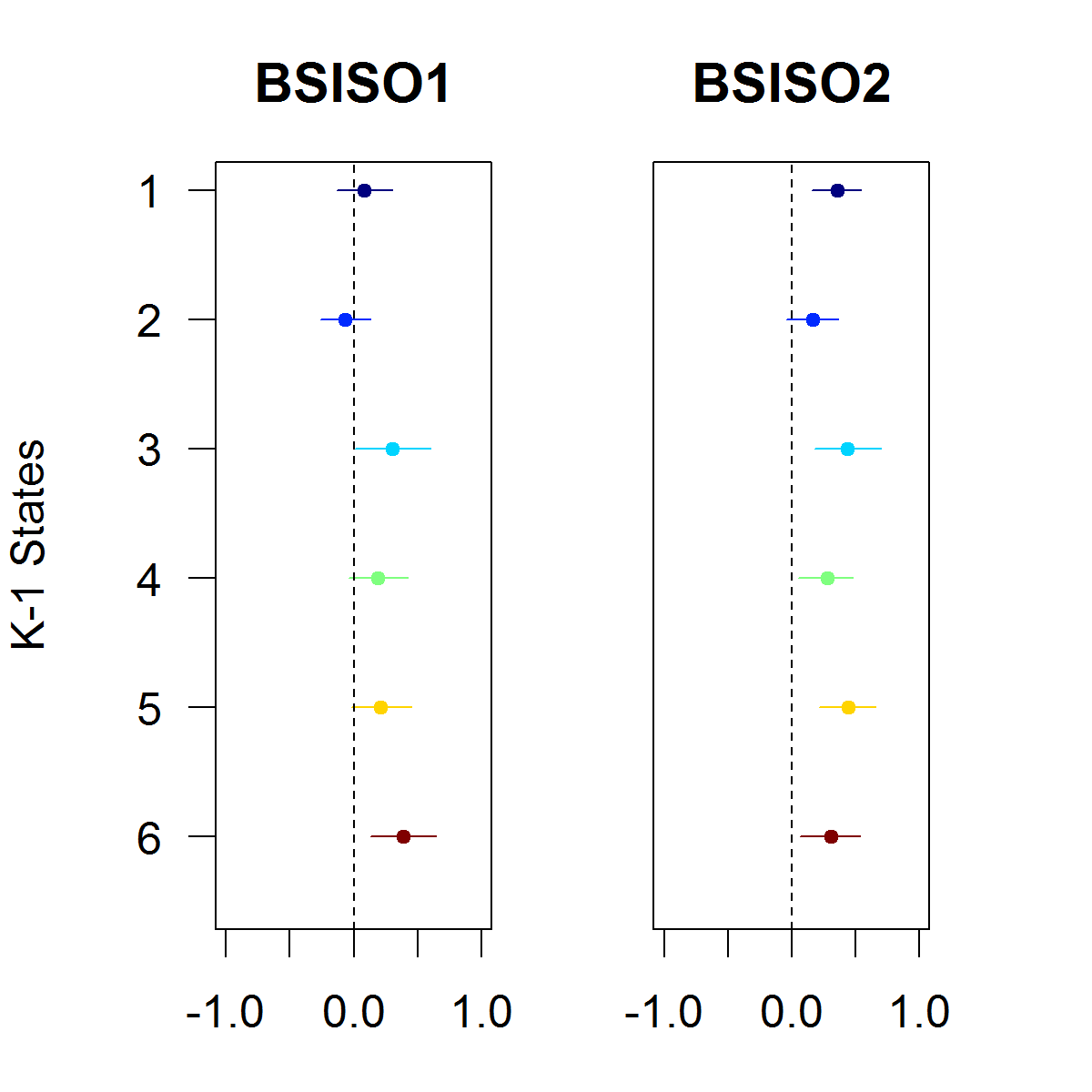}
\end{center}
\caption{Coefficients for the exogenous variables ${\bm x}$ influencing the transition probabilities for each of K-1 states. There is one dot for each of the K-1 states (the kth state is set to zero). The 95\% PI bands are given as a line around each dot.}
  \label{betaX}
\end{figure}

\begin{figure}[htp]
\begin{center}
\includegraphics[scale=0.45]{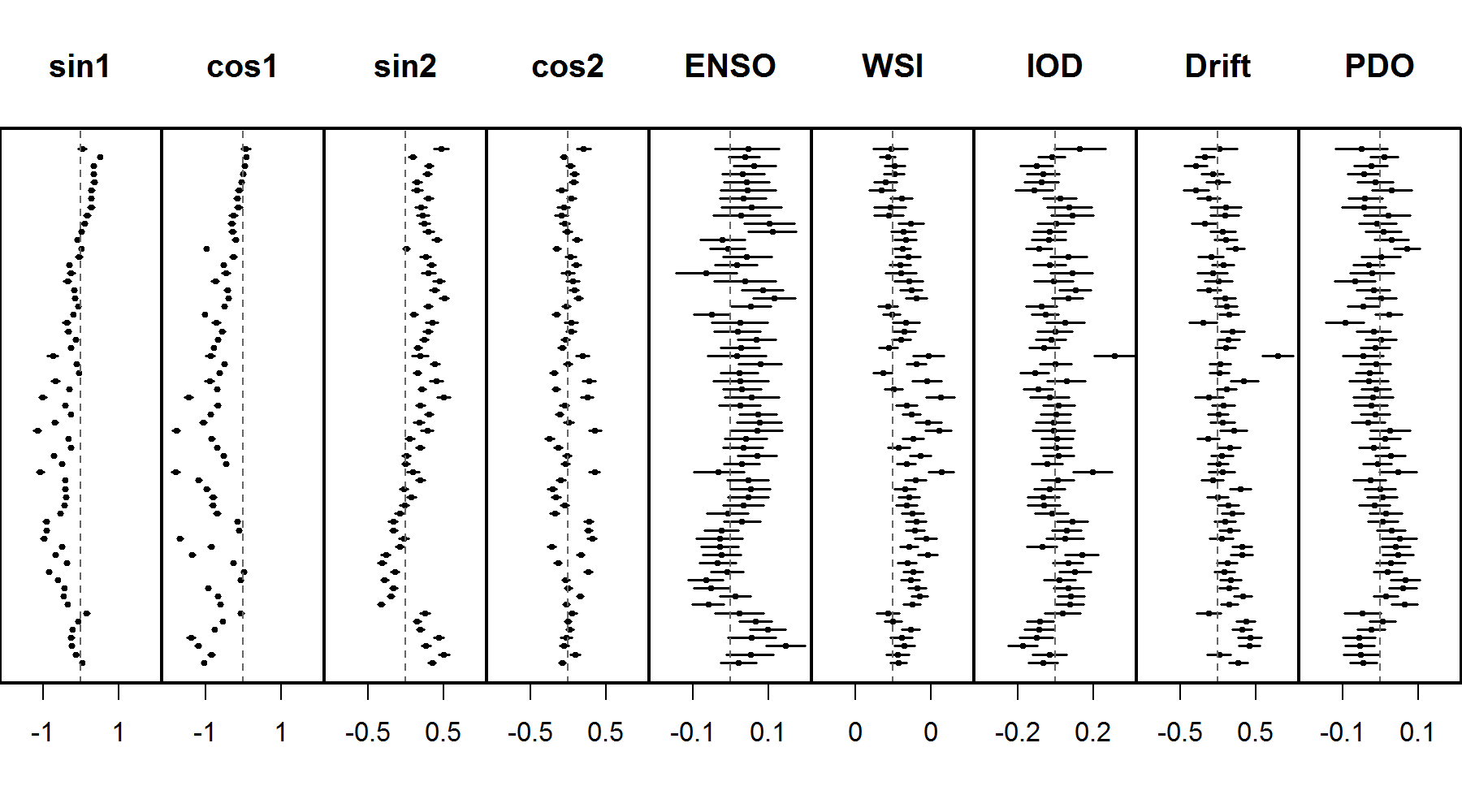}
\end{center}
\caption{Coefficients for the exogenous variables ${\bm w}$ influencing the emission distributions for each of J stations. Station 1 at the bottom through Station 63 at the top. The 95\% PI bands are given as a line around each dot.}
  \label{betaW}
\end{figure}

As described in Section 4.2, several covariates are included in the model. The BSISO1 and BSISO2 variables in the ${\bf x}$ vector impact the hidden state evolution on the daily time scale, and the ENSO, WSI, IOD, drift, and PDO in the ${\bf w}$ vector impact the mixing weights of the emission distributions on the monthly scale. There are $K-1$ coefficients for the transition probabilities and $J$ coefficients for the emission distributions (one for each station). Figures \ref{betaX} and \ref{betaW} show the inferred values of the coefficients for the exogenous variables ${\bf x}$ and ${\bf w}$ respectively, and their 95\% PIs (for the last 4000 draws from the posterior to ensure full convergence had happened, see \cite{supplementA} for trace plots); they are considered to be statistically significant if the PIs do not contain zero (vertical dashed line). The Bayesian approach has made this type of significance testing of the exogenous variables possible, many other NHMM algorithms only find point estimates for parameters of interest.    

\begin{figure}[htp]
\begin{center}
\includegraphics[scale=0.48]{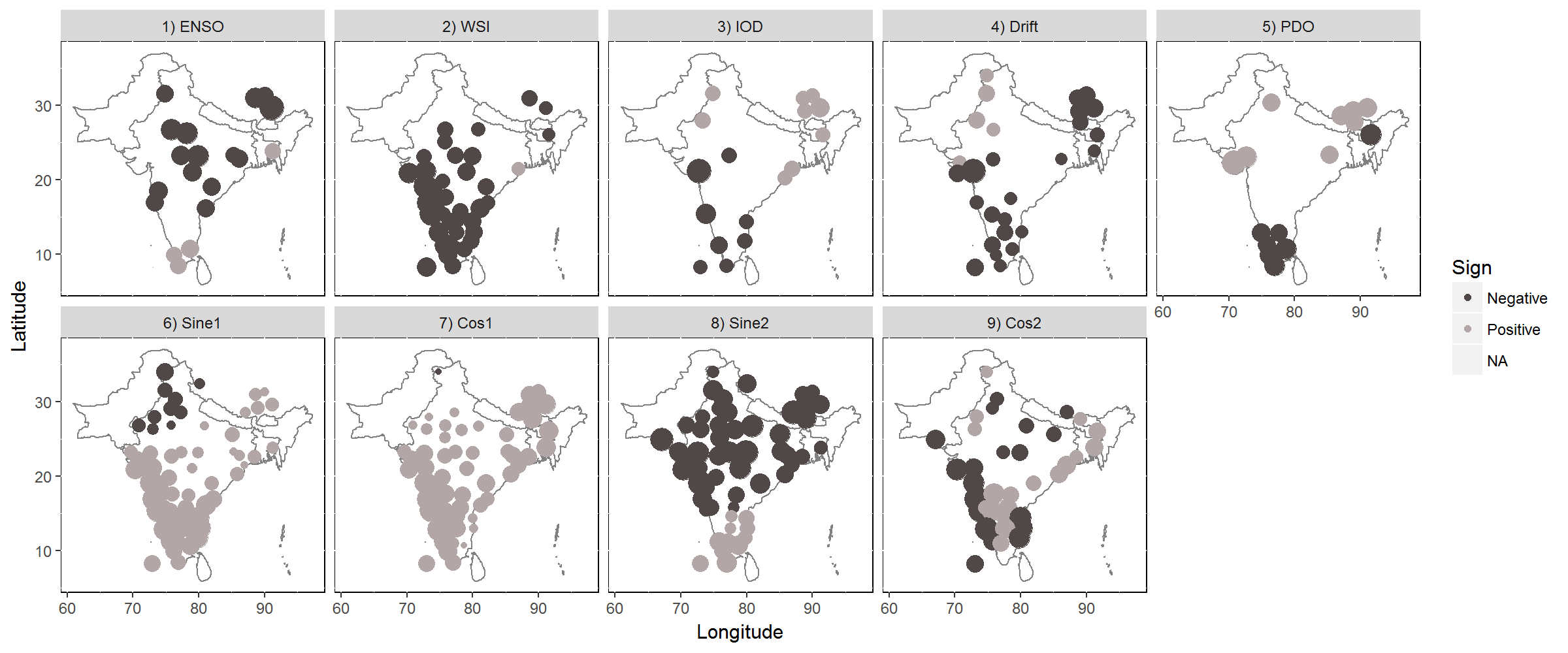}
\end{center}
\caption{Coefficients for the exogenous variables ${\bm w}$ influencing the emission distributions for each of J stations. 95\% PI for the coefficients having positive values (dark) and negative values (light) (coefficients containing zeros omitted); the magnitude of the coefficient is given by the relative size.}
  \label{betaWmap}
\end{figure}

Figure \ref{betaX} shows the coefficients for the exogenous variables
of $\bm x$, which affect the transition probabilities of the hidden
state evolution. Of the $K-1$ coefficients for each exogenous variable
at least one is well away from zero in each set. In the case of
  BSISO 2, all the coefficients are statistically significantly, as their 95\% PIs do not contain zero. The four seasonal harmonic input coefficients are not shown but were also all significant. Figure \ref{betaW} shows the coefficients for the exogenous variables ($\bm w$) for the emission distribution, one for each of the 63 stations (Station 1 at the bottom of the Figure through Station 63 at the top.) The harmonic terms representing rainfall seasonality are highly significant for most stations, consistent with Figure \ref{fig16}. Most of the other exogenous variables are significant at least at several stations, although their impacts are understandably much weaker than the seasonal modulation. Figure \ref{betaWmap} shows the mean of the coefficient values for the climate covariates, plotted geographically to highlight any spatial coherency and regionality in the relationships (note that the coefficient magnitudes depend on the scale of the covariate and are thus not comparable between panels). There is some indication that certain subregions are affected preferentially by particular exogenous variables, with the circulation index WSI showing the broadest scale impact. This is consistent with the direct physical relationship between monsoon rainfall and winds, while the remote climate modes (ENSO, IOD, PDO) have weaker impacts \citep{gadgil03a}.

\begin{figure}[htp]
\begin{center}
\includegraphics[scale=0.48]{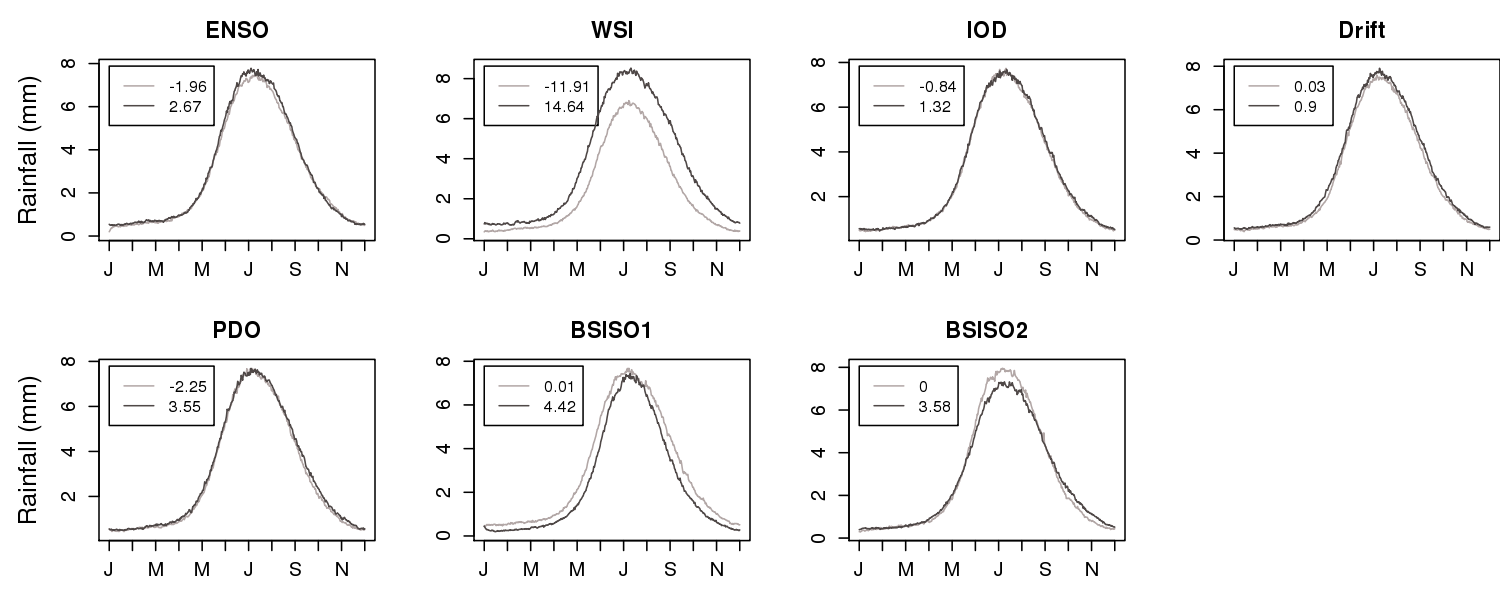}  
\end{center}
\caption{Maximum (dark) and minimum (light) for ENSO, WSI, IOD, drift, PDO, BSISO1, BSISO2 for each day of the year versus mean rainfall over 1000 simulated chains averaged over all stations (see \cite{supplementA} for individual stations).}
  \label{ENSOWSI}
\end{figure}

Figure \ref{ENSOWSI} shows mean rainfall amount over 1000 simulated chains versus day of the year, for the minimum (light) and maximum value (dark) of each of the exogenous variables (with all other inputs held at their mean values). The figure shows little difference in mean rainfall for the minimum and maximum values of the ENSO, PDO, and IOD covariates, when averaged over all stations. However, there is a marked difference in the minimum and maximum WSI value on the average rainfall amount, consistent with the broad scale geographical impact seen in Figure \ref{betaWmap}. There is a slightly longer and heavier monsoon when the drift term is higher,  indicating an upward trend in rainfall over the 27 year period.  BSISO1 and BSISO2 amplitudes are the only  inputs prescribed on a daily basis (whereas the other inputs are monthly). Smaller BSISO1 amplitudes are seen to be associated with longer and heavier monsoon seasons, while smaller BSISO2  is associated with heavier monsoon seasons but of the same duration. Thus, the monsoon tends to be stronger when the intraseasonal oscillation is less active which is physically consistent with fewer dry monsoon ``breaks'' in those years.  

\begin{figure}[htp]
\begin{center}
\subfigure[Station 3]{\label{a3}\includegraphics[scale=0.48]{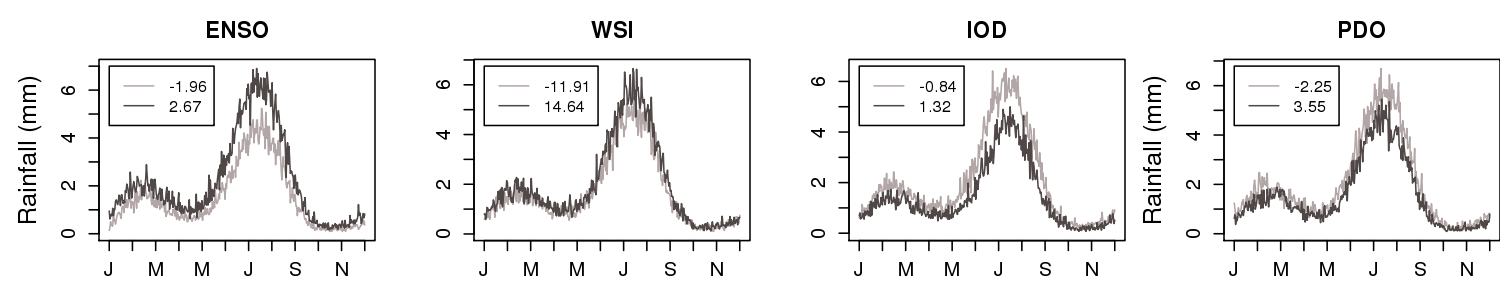}}
\subfigure[Station 40]{\label{a40}\includegraphics[scale=0.48]{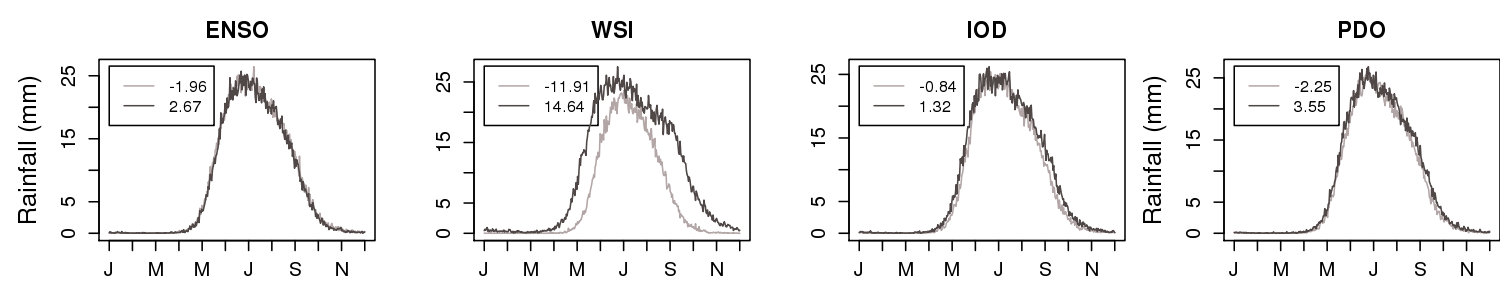}}\\
\subfigure[Station 52]{\label{a51}\includegraphics[scale=0.48]{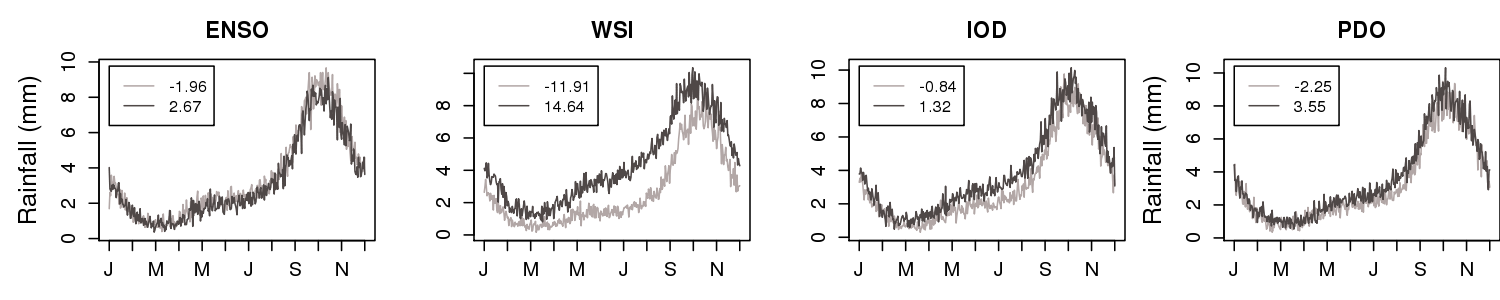}}\\
\end{center}
\caption{Three specific stations: maximum (dark) and minimum (light) for ENSO, WSI, IOD, PDO for each day of the year versus mean rainfall over 1000 simulated chains.}
  \label{inputs3}
\end{figure}

Figure \ref{inputs3} shows similar 1000-chain averaged annual simulations as Figure \ref{ENSOWSI}, but for the three selected stations. Years with strong monsoonal wind shear anomalies (WSI) are associated with a much longer summer monsoon rainfall season at the very wet station (station 40) on the west coast, while the impact is on peak rainfall at the ``dry'' station in NW India (station 3), not duration of the season. The SE India station (station 52), while nominally in the fetch of the NE monsoon that peaks in autumn, nonetheless also feels the summer SW monsoon as well when the monsoonal circulation (given by WSI) is strong, resulting in an extended rainfall season from May--Jan. The indirect climate covariates again have smaller impacts, though they can be quite large at the individual stations. Their impacts are large during summer at station 3 over over inland NW India, although physical interpretation is not straightforward. (See \cite{supplementA} for additional stations and exogenous variable plots.)


\subsection{Spatial Modeling}\label{spatialsec}
Finally, we assess the model's ability to reproduce the observed spatial correlations of the rainfall patterns. Stations that are closer together tend naturally to have more highly correlated measurements, although this can be modulated somewhat by local topography. The two-way station correlations of daily rainfall, for amount and occurrence, range between (0.04,0.41) and (0.04,0.77) respectively. We use two measures of spatial correlation between each pair of stations, the log odds ratio for occurrence and the Spearman's rank correlation coefficient for rainfall amount. The log odds ratio is calculated for the occurrence (binary classification) as the log of the number of matched days between the two chains divided by the number of differences \citep{hughes1999non}.

\begin{figure}[htp]
\begin{center}
\subfigure[NHMM: Rainfall amount]{\label{sa}\includegraphics[scale=0.35]{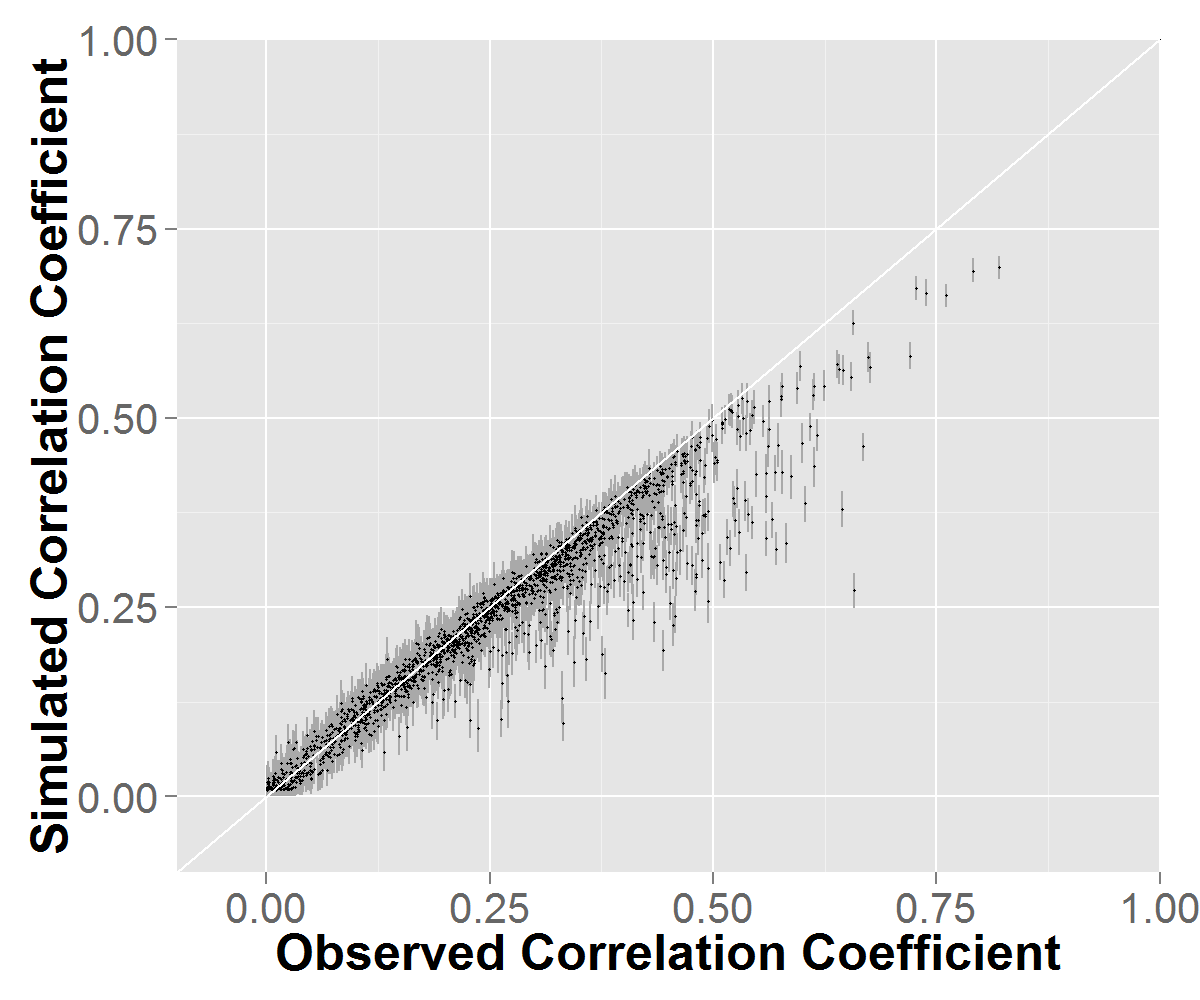}}
\subfigure[NHMM: Rainfall occurrence]{\label{so}\includegraphics[scale=0.35]{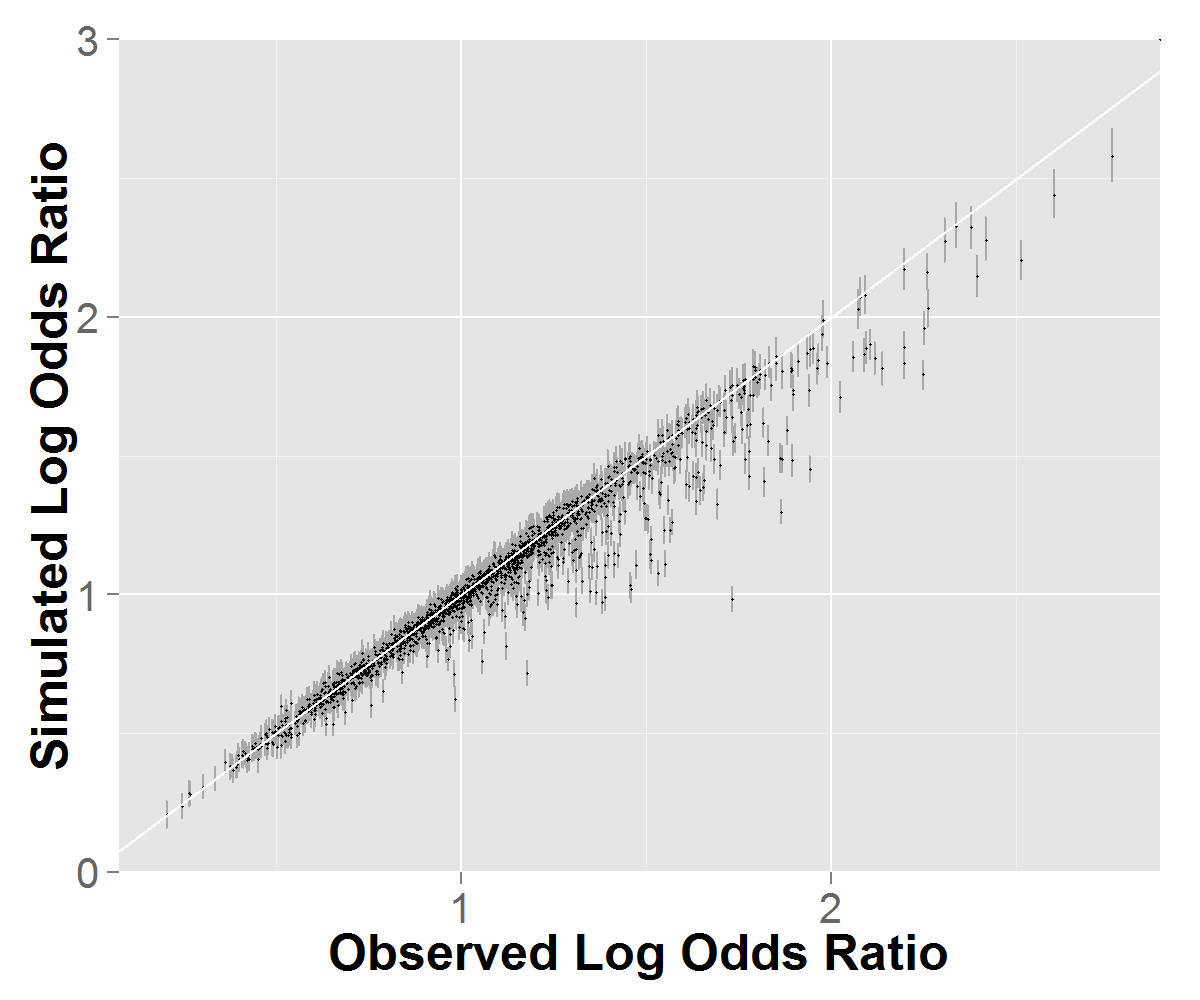}}\\
\subfigure[GLM: Rainfall amount]{\label{sa1}\includegraphics[scale=0.35]{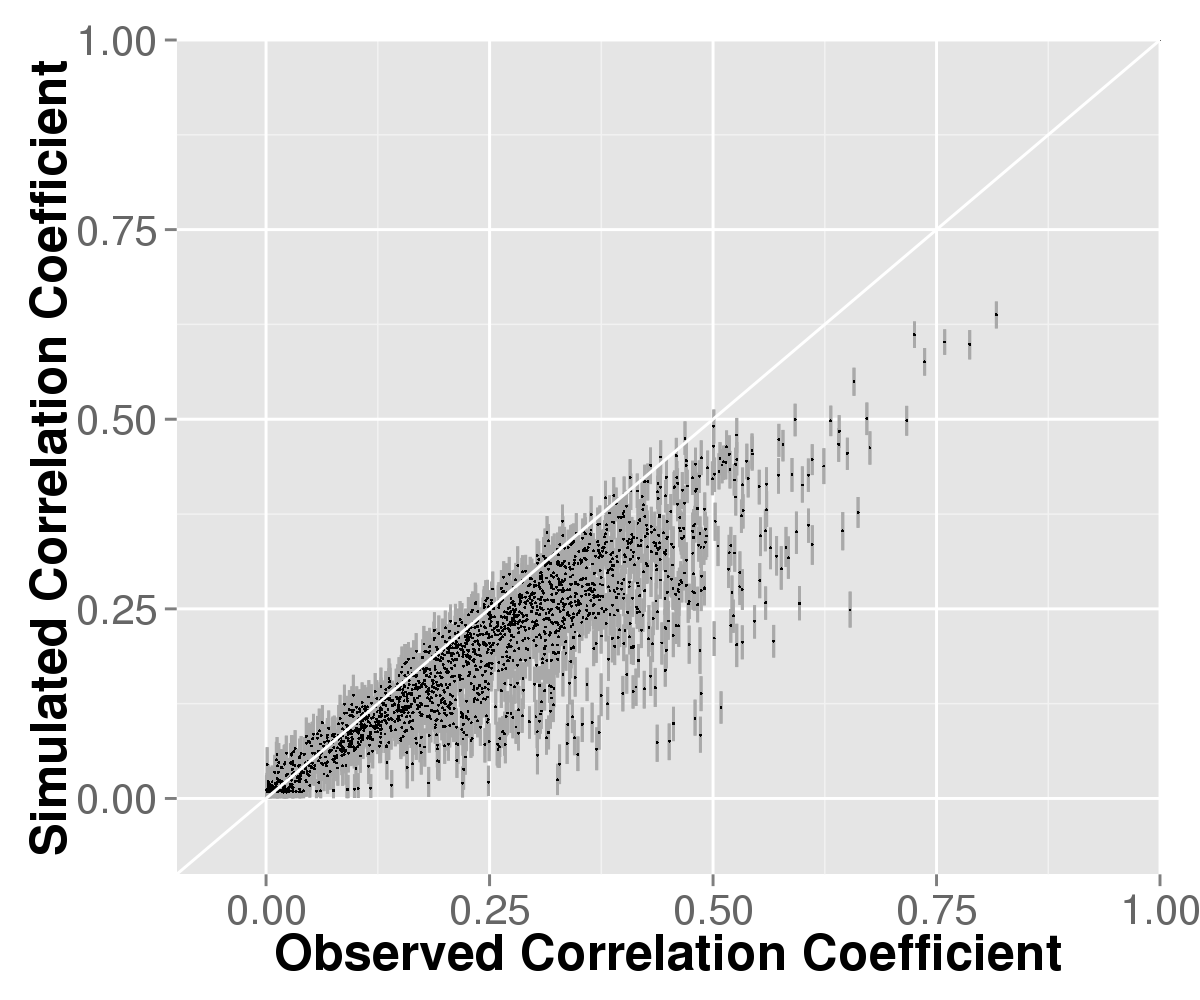}}
\subfigure[GLM: Rainfall occurrence]{\label{so1}\includegraphics[scale=0.35]{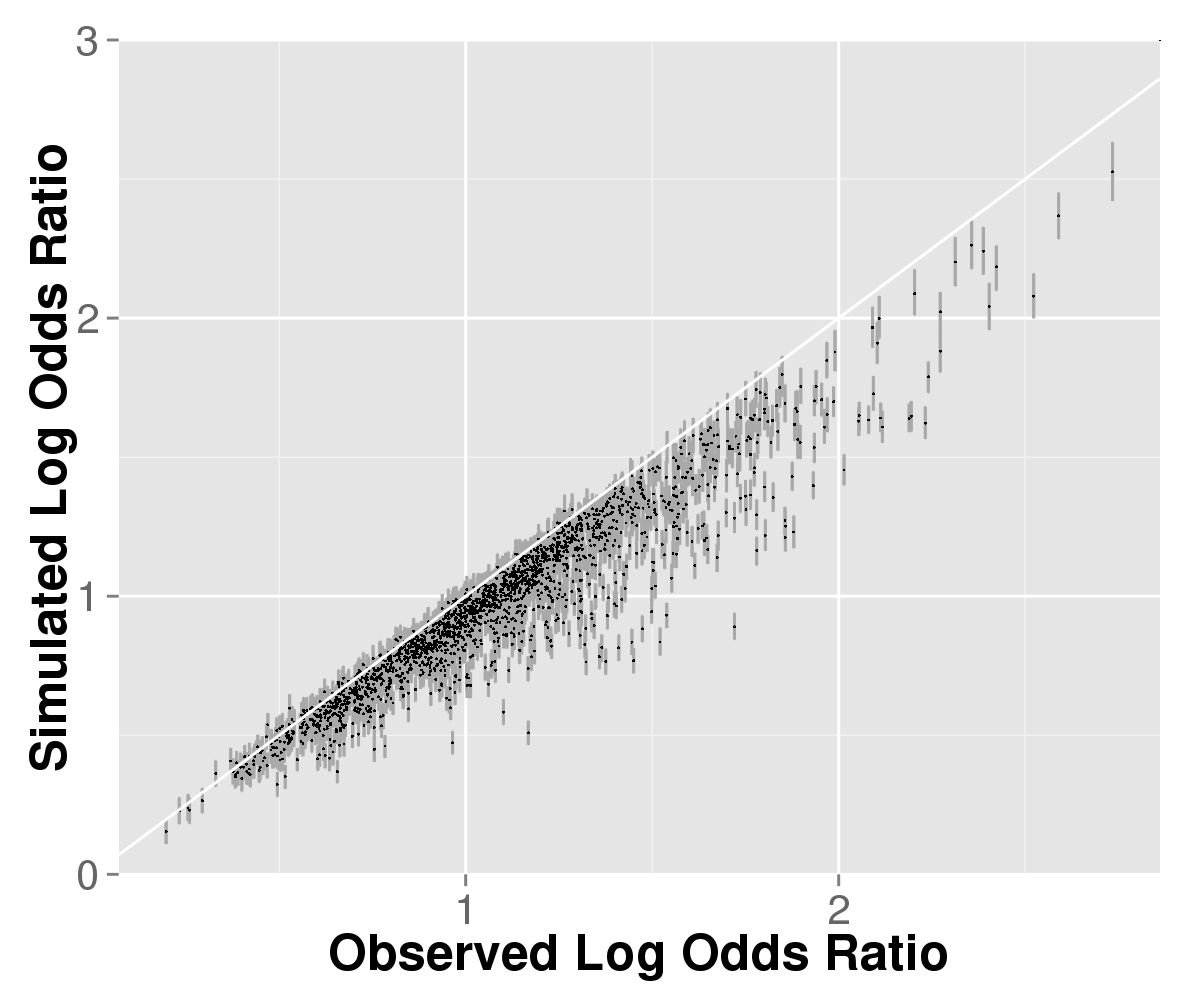}}
\end{center}
\caption{Pairwise station spatial correlation for the observed (x-axis) and the 1000 simulated chains (y-axis). The dot is the mean of the 1000 simulated chains and 95\% PI bands are in gray. Top: NHMM with state structure. Bottom: GLM model with no state structure. }
  \label{rain-spat}
\end{figure}

We compare the observed empirical pairwise correlations to correlations obtained from the 1000 simulated chains of daily rainfall of the same length (each 27 years in length) in Figure \ref{rain-spat}. The x-axis corresponds to the observed  correlations  and the y-axis to the simulated correlations. 95\% PI bands are included for the simulations. If the model were able to fully reflect the observed spatial correlations then the points in
 the figure should lie around the diagonal line. The upper panel shows that the spatial correlations tend to 
 be systematically underestimated by the NHMM, which is a known issue in rainfall 
 modeling when assuming that the station variables are conditionally independent given 
 the state variable \citep{hughes1999non, kirshner05, germain10}. For comparison, the spatial correlations from an equivalent GLM model (with no state structure, which includes the ${\bf w}$ variables, but not ${\bf x}$ variables, as described in Appendix A as Model 3) are shown in the lower panel. The simulated GLM correlations are less accurate than those of the NHMM, indicating that the state variables are contributing to better modeling of spatial dependence. Further improvements
 could be made by going beyond the conditional independence assumptions within each state,
 e.g., by using the type of tree-structured station dependence developed in \cite{Kirsh04}. (See \cite{supplementA} for similar plots for the three held out years of data.) For more analysis of isotropy and correlation see \cite{supplementA}; these plots show that the NHMM is capturing most of the correlation from the data. Future work could include adding more complex spatial structure to the model but this would have to consider the computational cost as some changes would be prohibitive.

\section{Conclusion}\label{sec:conc}

We described a Bayesian implementation of the NHMM based on the Polya-Gamma latent
variable scheme, allowing for analysis of larger multivariate data sets than possible with prior approaches.  The model allows for exogenous variables to influence both the transition probabilities of the hidden states and the emission distributions. The overall approach is flexible in that it can handle non-normally distributed multiple time series of daily data such as rainfall. Sampling is done through a data augmentation approach which removes the need for tuning parameters and makes it nearly automatic.  

We illustrated how the framework allows fitting  a
multivariate NHMM  for daily rainfall simulation allowing for  incorporation of covariate information of different forms, which is particularly attractive for downscaling of global climate model predictions and projections.
In particular we applied the approach to modeling of rainfall data  over a large historical collection of daily weather  station data across the Indian region. The general distributional properties of Indian rainfall, including seasonality,  were shown to be well captured by the model, and spatial correlations between stations adequately captured by the hidden states. The model was shown to be provide particular meteorological insight into the roles of  monsoon wind shear strength and intraseasonal wave activity on the seasonality of rainfall in the Indian region.
In particular, it enables a novel analysis of rainfall variability integrated from daily-to-seasonal time scales and from local-to-subcontinental spatial scales. This complements the methods traditionally used in monsoon diagnostic and predictability studies that often focus on correlation analysis between variables for a particular spatio-temporal scale, such as the predictability of seasonal averages of all-India rainfall, e.g., \citep{shukla}. More broadly, the Bayesian framework  provides uncertainty estimates for all parameters of the model, allowing for assessment of the impact of each exogenous climate variable. The NHMM can be used to generate predictive chains and chains with different levels of the exogenous variables, providing both chains of the hidden states and emission distributions that can be used for model comparison and conditional simulation.

Future work could be aimed at adding a more complex spatial structure between stations to this Polya-Gamma NHMM. Or the conditional independence assumption could be relaxed and subregions could be considered each with their own states. On smaller data sets, some have allowed $K$ to be estimated, this could be added to the model. We could also change the modeling assumptions to estimate a coefficient for each of the exogenous variables for each of the states (not the same coefficient for all states). Other NHMM assumptions like having individual coefficients for the transition probabilities could be included $\rho_{ij}$ instead of just $\rho_j$ in Equation \ref{eq1}. All changes of this type would include further investigation to model complexity and computational expense as some changes would introduce prohibitive.




\section*{Acknowledgment}
The rainfall data set was obtained from the Climate Prediction Center, National Centers for Environmental Prediction, National Weather Service, NOAA, U.S. Department of Commerce, from the Research Data Archive at the National Center for Atmospheric Research, Computational and Information Systems Laboratory. http://rda.ucar.edu/datasets/ds512.0. This research was supported by the U.S. Department of Energy, Office of Science, grant DE-SC0006616.

\break


 \section*{Appendix A: Model Selection}\label{sec:modelselection}
 
\subsection*{Model Selection Criteria}\label{mc}
Model selection has two purposes: to choose the number of states ($K$) and select the variables in ${\bf x}_t$ and ${\bf w}_{t,s}$. The number of hidden states needs to be determined for the model. More flexible models that allow for change in dimension tend to have complex algorithms that are computationally expensive (i.e., reversible-jump MCMC) \citep{green1995, rr00, meligkotsidou2011forecasting}. With such large data sets, we can run the NHMM with a few values of $K$ (number of hidden states) and use a model selection criterion to choose an optimal number of states. The other model choice is the selection of meaningful exogenous variables. Many exogenous variables can be included in ${\bm w}$ and ${\bm x}$ but some may not be statistically significant. 
 
There are two types of metrics we can consider for the model choice decision: model fit metrics (in-sample) or predictive metrics (out-of-sample). Standard in-sample model fit metrics like Akaike information criterion (AIC), Bayesian information criterion (BIC), and deviance information criterion (DIC) account for the number of parameters used in the model \citep{AIC, schwarz78,spiegelhalter2002}. We found that in this model because some of the assumptions of the DIC failed to hold, that it tended to choose over-parametrized models. Bayes factors (BF) are not considered because the model has non-informative priors on many of the parameters \citep{kass95, dempster97}. Following the recursive method given in \cite{scott02} to calculate the log-likelihood, we report the BIC values which are calculated as negative two times the log-likelihood plus a penalty for the number of parameters times the log number of observations (the parameters are counted: S=63, A=9, B=6, K*(K-1) for ${\bm \xi}$, B*(K-1) for ${\bm \rho}$, K*S for ${\bm \beta}_0$,  A*S for ${\bm \beta}_1$, (K-2)*S for the ${\bm \gamma}$ cutpoints, 2*S*K for ${\bm \lambda}$, and ${\bf z}$, ${\bf p}$, ${\bf L}$, ${\bf M}$ are latent variables that integrate out and are not counted). The second type of model metric is a predictive measure (out-of-sample) and we consider it the preferable method in this situation. We plan to hold out the last three years of the time series and perform a predictive log score (PLS) \citep{gneiting2007strictly,meligkotsidou2011forecasting}. After the last observation $y_{T,s}$, there are $r=1,\ldots,R$ predictive time steps to the simulated chain ($R=3*365$ for three years of predictive chains); the PLS is given by: $\sum_{r=1}^R \log \left(1/N \sum_{n=1}^N \prod_{s=1}^S f(y^*_{T+r,s})\right)$ where $N$ is the number of MCMC iterations.

The PLS is calculated for the predictive ability of the model and the BIC is calculated for the model fit. The BIC is calculated from the 27 years model fit (p is the parameter count of the model) and the PLS compares held out three years to predictive conditional chains of three years generated from the model, see Table \ref{Table-rainK1} and Table \ref{sm-table} for results. Table \ref{sm-table} shows several models for comparison with the NHMM and some different configurations of the exogenous variables. Table \ref{Table-rainK1} shows several options of the number of hidden state for the preferred model from Table \ref{sm-table}. Overall, the hidden states describe general spatial rainfall patterns, some states capture drier days while others describe wetter weather patterns. This large region may require more states (e.g., seven to twelve) than a more local region, where three to six states might suffice.

\subsubsection*{Model Selection Results}
We run a few baseline models to compare with the NHMM, see Table \ref{sm-table}. Model 1 is set up with no states ($K=0$) and no exogenous variables; this spreads rainfall homogeneously throughout the year. Model 2 is a standard weather state NHMM with all ${\bf x}$ inputs (i.e., BSISO1, BSISO2, seasonality (four harmonic components)); the optimal number of states for this model is $K=8$ using BIC. Model 3 treats each station independently ($K=0$) with the GLM linking inputs in ${\bf w}$ (i.e., seasonality (four harmonic components), ENSO,WSI, IOD,  Drift, PDO) with the mixing weights (there are no hidden states, thus no ${\bf x}$ inputs). This model is similar to \cite{katzparlange95, furrer07, ailliot12}, where only a single station is modeled.  Model 4 uses all inputs of ${\bf x}$ and ${\bf w}$; the placement of the inputs into either ${\bf x}$ or ${\bf w}$ was chosen by the climate scientists based on physical properties of the variables (i.e., larger regional variables and shorter time scale in ${\bf x}$). Other combinations and inputs were tested but these were the ones that were significant to the model. One of the other models tested was one with only ${\bf x}$ inputs, but it did not perform as well as models including ${\bf w}$.

Two other models were also considered that had similar (slightly worse) BIC and PLS scores than Model 4. One was a GLM-HMM \citep{holsclaw2015bayesian, heaps15} which includes all possible exogenous variables in ${\bf w}$. This model performed similarly in metrics but has some limitations. Because of climate change, it is of interest to forecast daily rainfall based on evolving ${\bf x}$ variables to modulate the hidden state distributions; the GLM-HMM type of model is stationary and has no mechanism for forecasting climate change like the NHMM. Additionally, another model was considered where all exogenous variables were included in both ${\bf x}$ and ${\bf w}$. This model performed similarly in metric scores to Model 4 (where variables were limited to being in either ${\bf x}$ or ${\bf w}$) but this model had far more parameters and suffered from lack of physical interpretability as the coefficients of ${\bf x}$ and ${\bf w}$ were highly correlated. The most parsimonious and interpretability model is the NHMM with exogenous variables each included once, either in ${\bf x}$ or ${\bf w}$ based on their physical characteristics. Each algorithm was run 2000 iterations with an additional 10\% burn in; the samples of the parameters converged quickly to stationarity and the samples mixed well with no thinning (see \cite{supplementA} for run times and trace plots). The final model was run 10000 iterations with and additional 20\% burn in period.

\begin{table}
\begin{center}
\caption{Comparing models for rainfall (*indicated the best value).}
\begin{tabular}{|clc|ccc|}
  \hline
No. & Model & States& p& BIC &PLS\\	
	\hline
1.&  No inputs (no {\bf x} or {\bf w})         & $K=0$& 252 & 1565454 & -85.7 \\
2.&  NHMM for {\bf x} (no {\bf w})             & $K=8$& 2079& 1360954 & -74.2 \\
3.&  Indep. GLM with {\bf w} (no {\bf x})      & $K=0$& 819 & 1406455 & -77.0 \\
4.&  NHMM with {\bf x} and {\bf w} partial     & $K=7$& 2283& 1344892*& -73.6*\\
   \hline
 \end{tabular}
    \label{sm-table}
\end{center}
\end{table}

Model 4 had preferable BIC and PLS metrics over all other models. Table \ref{Table-rainK1} shows the the metrics for choosing the number of hidden states ($K$) for this model (other models had similar values of $K$). The table also shows the number of parameters (p); for parsimony we want to choose a model with maximum PLS, minimum BIC, and minimum p (* denotes these values on the Table).  $K=1$ denote a model with a single constant state (which is equivalent as having no states $K=0$). Table \ref{Table-rainK1} shows that the BIC achieves local minima around seven to ten states. The PLS continues to improve with increased number of states. For parsimony it is sometimes best to choose the number of states where the PLS value is no longer improving as rapidly, this also happens around seven to ten states. 

We compare the PLS scores of the NHMM (Model 4) to a baseline model with no states (Model 3). Model 3 fit independent GLM models to each station, whereas Model 4 includes the hidden states and spatial information. Model 3 has a PPL of -77.0 for the baseline model compared to the NHMM with $K=7$ states with a PPL of -73.6. The difference between the two log scores over the three forecast years (predictive conditional chains) is given by $\exp((-73.6-(-77.0))/3)=3.1$, which means the NHMM is 3.1 times better at annual predictive ability. Also, we compare the PLS for Model 4 with $K=7$ states and $K=15$ states: $\exp((-73.0-(-73.6))/3)=1.2$ and find only a 1.2 times better annual predictive ability by including the eight additional states.

\begin{table}
\begin{center}
\caption{Choosing the number of states for the rainfall model. X and W have inputs based on physical properties of the region (*desirable numerical score).}
\begin{tabular}{|l|ccc|}
  \hline
K & p & BIC & PLS  \\ 
  \hline
0-1  & 693* & 1405643 & -77.1 \\ 
  2  & 953  & 1376646 & -75.6 \\ 
  3  & 1215 & 1366881 & -75.0 \\ 
  4  & 1479 & 1356233 & -74.4 \\ 
  5  & 1745 & 1351150 & -74.0 \\ 
  6  & 2013 & 1346841 & -74.0 \\ 
  7  & 2283 & 1344892 & -73.6 \\ 
  8  & 2555 & 1342660 & -73.6 \\ 
  9  & 2829 & 1342309 & -73.5 \\ 
  10 & 3105 & 1341327* & -73.3 \\ 
  11 & 3383 & 1341566 & -73.2 \\ 
  12 & 3663 & 1341514 & -73.1* \\ 
  13 & 3945 & 1341798 & -73.2 \\ 
  14 & 4229 & 1342378 & -73.0*\\ 
  15 & 4515 & 1342853 & -73.0* \\ 
   \hline
 \end{tabular}
    \label{Table-rainK1}
\end{center}
\end{table}

 \section*{Appendix B: Climate Variable Details}\label{sec:climvar}
\TR{Six established climate indices related to rainfall in India are: Westerly wind Shear Index (WSI), El Ni\~no/Southern Oscillation (ENSO), Indian Ocean Dipole (IOD), Pacific Decadal Oscillation (PDO), and two components of the boreal summer intraseasonal oscillation (BSISO1 and BSISO2).
\begin{enumerate}
	\item [1] WSI: Year-to-year (interannual) changes in the strength of the summer monsoon winds are closely linked with monsoon rainfall variations, and we use the Westerly Shear Index (WSI), as defined in \cite{wang99} as WSI1, to represent these. The WSI is defined by the vertical shear of the zonal wind (u850 −- u200 ) averaged over the box (5N--20N, 40E--80E), and was used in an NHMM for Indian rainfall by \cite{greene11}. We use monthly-averaged values, with the mean seasonal cycle subtracted, so as to focus on interannual variations in the monsoon circulation. 
\item [2-3] ENSO and IOD: ENSO and IOD indices were computed from the NOAA Extended Reconstructed Sea Surface Temperature Dataset, version 3b \cite{SmithEtAl_JCL08}, via the IRI Data Library (\url{http://iri.columbia.edu}). The El Ni\~no/Southern Oscillation (ENSO) and Indian Ocean Dipole (IOD) are known influences on rainfall on interannual time scales \citep{gadgil03a}. Monthly sea surface temperature (SST) in the Nino3.4 region (150W-90W, 5N-5S) are used to define the ENSO index the monsoon tends to be stronger during the La Nina phase, when this ENSO index is \emph{negative} \citep{gadgil03a}. The IOD index is defined by the difference in monthly SST anomalies in the western (50E--70E, 10N--10S) and eastern (90E--110E and 0S--10S) equatorial Indian Ocean; the monsoon tends to be stronger when IOD is positive \citep{gadgil03a}.   
\item [4] PDO: While the Pacific Decadal Oscillation (PDO) has a less well understood impact \citep{JosephEtAl_AOSL13}. The PDO index is defined by \cite{ZhangEtAl_JCL97} to be the leading PC of monthly SST anomalies in the North Pacific Ocean, poleward of 20N. The monthly mean global average SST anomalies are removed to separate this pattern of variability from any ``global warming'' signal that may be present in the data. This data set set is from University of Washington (\url{http://research.jisao.washington.edu/data_sets/pdo/})
\item [5-6] BSISO 1 and 2: On sub-seasonal time scales Indian monsoon rainfall is impacted by the boreal summer intraseasonal oscillation (BSISO), data obtained from the APEC Climate Center (APCC, \url{http://www.apcc21.org}) \citep{LauChan_MWR86, YooRobertsonKang_JCL10}, for which we use the two indices BSISO 1 and 2 defined by \citep{LeeEtAl_CD13}.
\end{enumerate}
}

\TR{The cross-correlations between these daily series are given in Table \ref{Table-2cor} and are relatively low. Monsoon circulation anomalies (WSI) are quite strongly related to ENSO and PDO (r = -0.51 and 0.41 resp.), less strongly with IOD (r = 0.28), but not to the BSISO. }

\begin{table}
\begin{center}
\caption{Two-way Input Correlations}
\begin{tabular}{|l|cccccc|}
   \hline
 & ENSO & WSI & IOD & PDO & BSISO1 & BSISO2 \\ 
  \hline
ENSO & --- & -0.51 & 0.28 & 0.41 & -0.01 & 0.02 \\ 
  WSI &  & --- & -0.19 & -0.25 & -0.01 & -0.04 \\ 
  IOD &  &  & --- & 0.09 & -0.03 & -0.02 \\ 
  PDO &  &  &  & --- & -0.08 & 0.01 \\ 
  BSISO1 & &  &  &  & --- & 0.07 \\ 
  BSISO2 &  &  &  & &  & --- \\ 
   \hline
 \end{tabular}
    \label{Table-2cor}
\end{center}
\end{table}



\bibliographystyle{imsart-nameyear} 
\bibliography{lit-aoas}

\clearpage
\end{document}